\documentclass[10pt,conference]{IEEEtran}

\usepackage{amssymb}
\usepackage{amsfonts}
\usepackage{enumitem}
\usepackage{amsthm}
\usepackage{optidef}
\usepackage{caption}
\usepackage{subcaption}
\usepackage{bm}
\usepackage{bbm}
\usepackage{booktabs}
\usepackage{float}
\usepackage[ruled,vlined]{algorithm2e}
\usepackage{tablefootnote}
\usepackage{flushend}
\usepackage{xcolor}
\usepackage{mathtools}

\theoremstyle{definition}

\theoremstyle{remark}

\DeclareMathOperator*{\argmaxA}{arg\,max} 
\DeclareMathOperator*{\maxA}{max}

\DeclareMathOperator*{\nc}{\mid c \mid}

\SetCommentSty{mycommfont}

\usepackage{graphicx}
\graphicspath{Images/}
\usepackage{subfiles} 

\begin{document}

\title{SneakPeek: Data-Aware Model Selection and Scheduling for Inference Serving on the Edge}

\author{\IEEEauthorblockN{Joel Wolfrath, 
Daniel Frink, and Abhishek Chandra}
\IEEEauthorblockA{Department of Computer Science and Engineering\\
University of Minnesota,
Minneapolis, USA\\
Email: \{wolfr046, frink012, chandra\}@umn.edu\\}
}

\maketitle

\begin{abstract}
Modern applications increasingly rely on inference serving systems to provide low-latency insights with a diverse set of machine learning models.
Existing systems often utilize resource elasticity to scale with demand.
However, many applications cannot rely on hardware scaling when deployed at the edge or other resource-constrained environments.
In this work, we propose a model selection and scheduling algorithm that implements accuracy scaling to increase efficiency for these more constrained deployments.
We show that existing schedulers that make decisions using profiled model accuracy are biased toward the label distribution present in the test dataset.
To address this problem, we propose using ML models--which we call SneakPeek models-- to dynamically adjust estimates of model accuracy, based on the underlying data.
Furthermore, we greedily incorporate inference batching into scheduling decisions to improve throughput and avoid the overhead of swapping models in and out of GPU memory.
Our approach employs a new notion of request priority, which navigates the trade-off between attaining high accuracy and satisfying deadlines.
Using data and models from three real-world applications, we show that our proposed approaches result in higher-utility schedules and higher accuracy inferences in these hardware-constrained environments.
\end{abstract}


\section{Introduction}

The widespread adoption of machine learning across a diverse set of applications has created a pressing need for efficient and scalable inference serving systems.
These systems provide APIs for executing pre-trained models and handle billions of requests daily at some companies~\cite{8327042}.
Inference serving systems often host a large number of model variants and must select an appropriate model variant to service each request, in addition to scheduling the request for execution.~\cite{infaas}.
Inference serving underpins several important applications, including healthcare monitoring~\cite{model-serving-health}, language translation~\cite{8327042}, recommendation systems~\cite{9773234}, and mobile applications~\cite{mobile-serve}.

While existing inference serving systems are often deployed in the public cloud, inference serving is also critical for applications that run in more constrained environments, such as edge clusters or private clouds.
Local deployments, such as those found in healthcare facilities or smart cities, are often necessary to avoid the increased cost, latency, and privacy concerns associated with streaming data over the wide-area network to a cloud setting.
However, there are several challenges associated with efficient inference serving in these local deployments.
First, such inference serving systems must service requests for \textit{multiple applications}, each with their own set of data streams which may require real-time processing. These applications also have unique service level objectives (SLOs), including low-latency deadlines which are critical to satisfy, especially for real-time applications~\cite{octopus,daisy}.
Next, local deployments often have resource constraints (e.g. may limited to a single GPU) and lack the resource elasticity of the cloud, which prevents systems from using autoscaling to meet demand. 
Finally, deep neural networks (DNNs) are a common model choice due to their high accuracy, but they can be computationally intensive.
An inference serving system may offer a collection of multiple DNN models for each application to choose from, with each model presenting a different latency-accuracy tradeoff (e.g., some models may be lightweight but less accurate, while others may have higher accuracy but take longer to execute)~\cite{daisy}. Thus, selecting and scheduling the right model for each request in a multi-application, constrained resource environment is challenging. 

Existing works largely focus on inference serving in public clouds or other settings which support hardware scaling~\cite{sponge, eurosys-mss, tolerance-tiers, rl-selection1, compress-sched, mark, infaas}.
A few recent works have proposed \textit{accuracy scaling} as an alternative approach when resource elasticity is unavailable~\cite{proteus,daisy,loki,jellyfish}.
The main idea is to estimate (profile) the accuracy of each model offline and dynamically trade-off between accuracy and throughput as system demand fluctuates.
This approach works in practice, but existing designs are suboptimal for two reasons:
(1) they are data-oblivious: profiled model accuracy is biased toward the test dataset and may not reflect the distribution of the out-of-sample data, thus reducing the quality of the associated scheduling decisions; and
(2) they consider each inference request in isolation: inference batching is only utilized opportunistically, rather than explicitly incorporated in the scheduling process, thus increasing the model loading overhead and reducing scheduling efficiency.

We show that existing systems that rely on profiled model accuracy make suboptimal model selection decisions when properties of the data are ignored.
We propose leveraging \textit{data-awareness} to understand how each model will perform over the actual data being processed in real-time.
In this framework, we treat class frequencies as parameters, which are dynamically estimated based on the data.
These parameters are then used to inform model selection decisions by providing sharper estimates of model accuracy.
We propose a joint model selection and scheduling algorithm for hardware-constrained, multi-application inference serving.
Our algorithm uses a new notion of request priority based on model accuracy and request deadlines, which attempts to maximize inference accuracy while minimizing deadline violations.
Furthermore, our approach greedily incorporates inference batching (or request grouping) into the resulting schedules. 
This grouping provides the scheduler with a more global view of request dependencies, while
reducing overheads and model swap latency in and out of GPU memory. 

\noindent
\textbf{Contributions.} We make the following research contributions:
\begin{itemize}[leftmargin=*]
    \item We formulate an optimization framework for joint model selection and request scheduling which implements accuracy scaling for hardware-constrained environments.
    \item We present our data-awareness mechanism, {\em SneakPeek}, which sharpens accuracy estimates based on the underlying data. SneakPeek is easily incorporated into existing schedulers and does not require modifications to DNN models.
    \item We propose heuristics for efficiently handling model selection and scheduling decisions in practice. These heuristics employ a new notion of priority as well as inference batching (grouping).
    \item We thoroughly evaluate our proposed approach using three, real-world applications and a variety of model types. Compared to the baselines, our methods achieve up to a 2x increase in \textit{utility}, a metric that combines model accuracy with a penalty for any missed deadlines.
\end{itemize}




\section{Preliminaries}
\label{sec:prelim}

\subsection{Motivating Applications}

Edge analytics is critical for smart city projects~\cite{killer-app}, as it processes data from a wide array of sensors and cameras to optimize traffic flow, manage waste, enhance public safety, and support urban planning. In the realm of surveillance, video analytics improves security by automatically identifying unusual behaviors or specific actions. This capability enables systems to alert security teams about potential threats, ensuring rapid, proactive responses~\cite{Olatunji2019}. Furthermore, modern surveillance extends beyond video alone, and requires capturing data and making real-time decisions in various sectors such as retail, transportation, and service industries.

Modern hospitals, assisted living facilities, and in-home care systems also collect and process real-time data from a variety of sources (such as wearable sensors, monitors, video cameras, etc.), which can be used to improve patient care and assist the medical staff~\cite{ieee-health-survey, Hassan2019}.
Many of these tasks require low-latency inference, since the results may directly affect patient well-being.
For example, care facilities may want to monitor patient movement or gait, which could alert medical staff if an elderly patient is wavering or falling over.
There are several other applications in this domain which can leverage inference serving, including arrhythmia detection\cite{ad1}, seizure forecasting~\cite{seizure1, seizure2}, and respiratory compromise~\cite{resp1,resp2}.
Since offloading data and tasks to a remote cloud is not an option due to latency, cost, and privacy concerns, many of these care facilities are equipped with edge clusters that have limited resource capacity and elasticity.
Real-time inferences can be generated in this setting by streaming  video frames or data from wearable sensors to the edge cloud and issuing recurring queries to monitor for critical events.
We use healthcare informatics as a running example in this work.

\subsection{System Model}


In our proposed inference serving framework, the system is tasked with selecting a model variant for each request and  scheduling it for execution.
Target applications wishing to access the inference-serving APIs first register their application with the system.
This registration process includes providing metadata for the application, uploading pre-trained ML models to be used for inference, and obtaining profiles for each of the models.
These profiles include estimates of inference latency, GPU memory usage, latency for loading the model into GPU memory, and model accuracy, which we assume is passed in as input by the application owner, but could also be collected by the system directly.
Users also indicate any new or existing data streams they wish to leverage, and can optionally specify recurring queries for their application, which defines how often requests should be issued to the system for each query.
Applications also have an associated service level objective (SLO) which specifies a deadline associated with each request.

\section{Problem Statement and System Architecture}
\label{sec:problem-and-arch}

\begin{figure}
    \centering
    \includegraphics[width=0.95\linewidth]{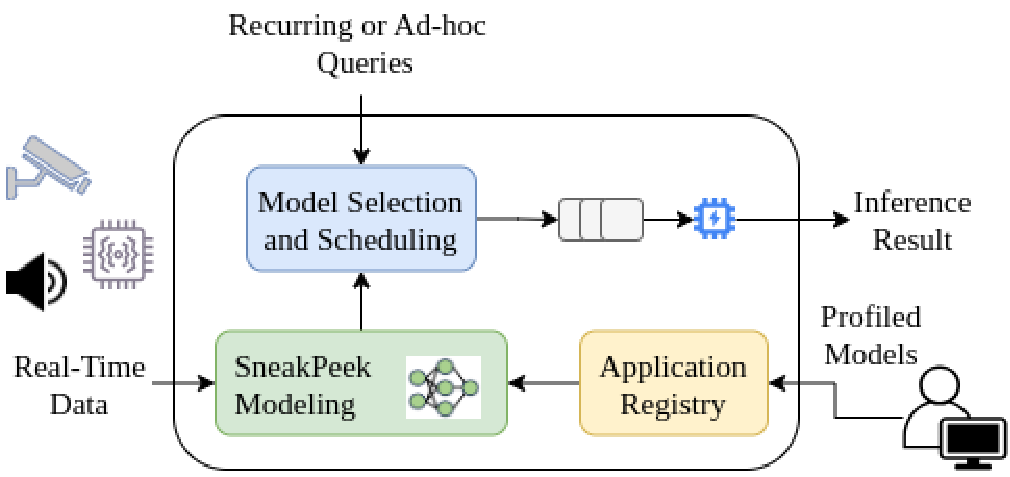}
    \caption{SneakPeek system model for inference serving}
    \label{fig:proposed-system}
\end{figure}

\subsection{Problem Statement}
Our system is tasked with providing inference serving in a hardware-constrained, multi-application setting.
In this paper, we focus on execution on a single GPU worker, but our techniques can be extended to multi-worker settings, as discussed in section \ref{sec:discussion}.
Each application registers one or more classification models\footnote{All of our approaches can be adapted to support other modeling types (e.g. regression) but we focus on classification to simplify the presentation.}, which the system can use to satisfy inference requests.
For a given set of requests, our task is to \textit{assign a model} to each request and \textit{schedule} the associated inference tasks for execution.
More formally, let $\mathcal{R}$ be our set of requests, where each request $r_i \in \mathcal{R}$ has an associated deadline of $d_i$.
Each request also belongs to a target application, which we denote $a_i$.
We define $\mathcal{M}_a$ to be the set of registered model variants for application $a$.

Our objective is to find a schedule $\mathcal{S}$ that maximizes the average \textit{utility} for all requests.
The utility of a request (defined below) represents the expected inference accuracy, which degrades if a deadline is missed. 
The schedule can be represented by a set of nonnegative integers, denoted $s_{ij}$, mapping requests to models. Here, a nonzero value of $s_{ij}$ indicates that request $r_i$ is assigned to model $m_j \in \mathcal{M}_{a_i}$ (for $i \in 1..|\mathcal{R}|$ and $j \in 1..|\mathcal{M}_{a_i}|$).
If $s_{ij}$ is positive, the integer value represents the execution order, relative to the other requests (with lower values of $s_{ij}$ being executed first, the first request being assigned an $s_{ij}$ value of 1).
We also define the execution start time $t_i$ for a given request $r_i$ (omitting the dependence on $\mathcal{S}$ for brevity):

\begin{equation}
    t_{i} = \sum_{j=1}^{|\mathcal{M}_{a_i}|} \sum_{h=1}^{|\mathcal{R}|} \sum_{l=1}^{|\mathcal{M}_{a_h}|} \mathbbm{1}_{s_{hl} < s_{ij}} \; \ell(m_l)
\end{equation}

\noindent
where $\ell(m_l)$ is the (profiled) inference latency for model $m_l$, \textit{including any context switch time required to swap the model variant into GPU memory}.
This works by finding all requests that are ordered ahead of request $r_i$ and summing their expected model latencies.

We now define the utility for a given request $r_i$:
\begin{equation}
    u_{a_i}(m_j, d_i, t_{i}) = Accuracy(m_j)[1 - \gamma_{a_i}(d_i, t_{i} + \ell(m_j))]
\end{equation}

\noindent
where $\gamma_{a_i}(d_i, e_i) \geq 0$ is a monotonically increasing 
penalty function, which generates positive values when a deadline is expected to miss (i.e. when the deadline $d_i$ is less than the expected request completion time $e_i$).
Some examples include a sigmoid function, linear penalty, or step function, e.g. $\gamma_{a_i}(d_i, e_i) = \mathbbm{1}_{d_i < e_i}$.
If all the deadlines are met, the utility is just the profiled accuracy\footnote{Note: Some implementations may prefer to normalize the accuracy values across applications, to ensure fairness.} of the selected model.
These utility (and penalty) functions are defined at the application level, and are the same for all requests belonging to an application.
Even applications which require high accuracy (e.g. in a healthcare setting), can define a utility function in a way which satisfies their requirements. For example, if we define our penalty function to be a constant, the optimization will strictly maximize accuracy.
We now define our optimization problem for scheduling and model selection, which seeks to maximize the average utility across all requests:

\begin{align}
\label{np-hard-opt}
\maxA_{\mathcal{S}} \quad & \frac{1}{|\mathcal{R}|} \sum_{i=1}^{|\mathcal{R}|} \sum_{j=1}^{|\mathcal{M}_{a_i}|} \mathbbm{1}_{s_{ij}>0} \; u_{a_i}(m_j, d_i, t_{i}) \\
\textrm{s.t.} 
\quad & \; \; s_{ij} \in \mathbb{Z}^{+}_0 \\
\quad & \; \; \sum_{j=1}^{|\mathcal{M}_{a_i}|} \mathbbm{1}_{s_{ij}>0} = 1, \; \; \forall i \\
\quad & \; \; s_{hl} \neq s_{ij}, \; \; \forall h,l \; \; \; \textrm{s.t.} \; \; \; h \neq i, \; s_{hl} > 0, \; s_{ij} > 0
\end{align}

\noindent
Constraint (4) forces the $s_{ij}$ terms to be non-negative integers.
A non-zero value indicates request $r_i$ is assigned model $m_{j} \in \mathcal{M}_{a_i}$.
The integer value of $s_{ij}$ indicates its order in the execution sequence.
Constraint (5) ensures that each request gets assigned exactly one model. 
Constraint (6) ensures that requests have distinct integers for ordering.
A simplified version of this problem is known to be NP-hard~\cite{springer-scheduling, siam-scheduling} (appendix \ref{appendix-hardness}).

Evaluating all candidate solutions is extremely expensive, with the number of solutions for $n$ requests (for a single application) totaling $n! | \mathcal{M}_{a_i} |^{n}$.
Computing an exact solution may be feasible when the number of requests and/or models is very small.
However, in many practical settings, we require an approximate solution that can be found quickly.

\subsection{System Architecture}
In our proposed system (figure \ref{fig:proposed-system}) application owners register their applications with the system and upload model variants than can be used to service their requests.
In addition, we assume model profiles are specified, which include accuracy measurements \textit{for every possible target label}, along with profiled latency measurements for the GPU.

After an application is registered, real-time data is then streamed to our SneakPeek module (section \ref{sec:sneakpeek}) which is a distinct process for asynchronous data staging, preprocessing, and sharpening accuracy estimates via data-awareness.
Inbound requests are enqueued during a scheduling window, then scheduled for inference.
The scheduler obtains metadata from the SneakPeek module, then assigns models and produces a schedule for the provided requests (section \ref{sec:scheduling}). These requests are then dispatched to the worker queue. 
The worker loads the data and model variant required to service the request and generates the inference result.

\section{SneakPeek: Incorporating Data-Awareness}
\label{sec:sneakpeek}

\subsection{Dependence on Model Accuracy}
Hardware-constrained scheduling algorithms in the literature often utilize accuracy scaling, \textit{which requires high-quality estimates of model accuracy for correctness.}
If the accuracy estimates for each model variant are poor, the resulting schedules will be suboptimal.
Existing schedulers are often \textit{data-oblivious} and rely on a single, profiled estimate of model accuracy for making decisions.
Relying on a single summary statistic to make decisions can cause schedulers (including exact solvers) to produce suboptimal solutions.

Model accuracy is not a static quantity; it often varies across target classes in a dataset~\cite{daisy}.
For example, consider the task of performing human action recognition over video frames.
Some actions, such as walking and sitting, are easier for a model to distinguish than others, such as loitering or talking on the phone. This makes the profiled model accuracy very sensitive to the frequency of each class in the test data set.
Therefore, it is an (often unstated) requirement that the profiled test set match the distribution of classes that will be observed out-of-sample.
However, even if a practitioner can match the distribution exactly, using a static accuracy value ignores the heterogeneity present in the data, which can be used to obtain better accuracy values for each inference request.

We can also observe this phenomenon by examining model accuracy analytically.
In the context of multi-class classification with a set of class labels $c$, the profiled accuracy of a model $m$ is computed as~\cite{acc-def}:

\begin{equation}
    \textrm{Accuracy}(m) = \frac{\textrm{tr}(Z)}{\sum_{i=1}^{\nc} \sum_{j=1}^{\nc} z_{ij}}
\end{equation}

\noindent
where $Z = [z_{ij}]_{1\leq i,j \leq \mid c \mid}$ is the confusion matrix generated by evaluating the model on a test data set. We can rewrite this expression as:
\begin{align}
     \textrm{Accuracy}(m)
     &= \sum_{i=1}^{\nc} \left( \frac{\sum_{j=1}^{\nc} z_{ij}}{\sum_{j=1}^{\nc} \sum_{k=1}^{\nc} z_{jk}} \right) \frac{z_{ii}}{\sum_{j=1}^{\nc} z_{ij}} \\
     &= \sum_{i=1}^{\nc} \; \theta_i \; \frac{z_{ii}}{\sum_{j=1}^{\nc} z_{ij}} \label{eq:theta-acc}
\end{align}

\noindent
where $\theta_i$ is the frequency of class $c_i$ in the test set and the remaining term is the recall for $c_i$.
While the recall for a given class depends on the trained model, \textit{the frequency of each class ($\theta$) depends exclusively on the test data set}. This implies that the profiled accuracy can be a poor estimator of out-of-sample accuracy if the class frequencies do not match the frequency in the test data set.

To address this issue, we propose dynamically computing model accuracy\footnote{In addition to accuracy, several other scoring rules can be rewritten in terms of $\theta$ (appendix \ref{appendix-scoring-rules}).} by treating $\theta$ as a parameter in equation \ref{eq:theta-acc} and estimating it using the data rather than implicitly assigning it the frequencies in the test set.

\begin{figure}
    \centering
    \includegraphics[width=0.99\linewidth]{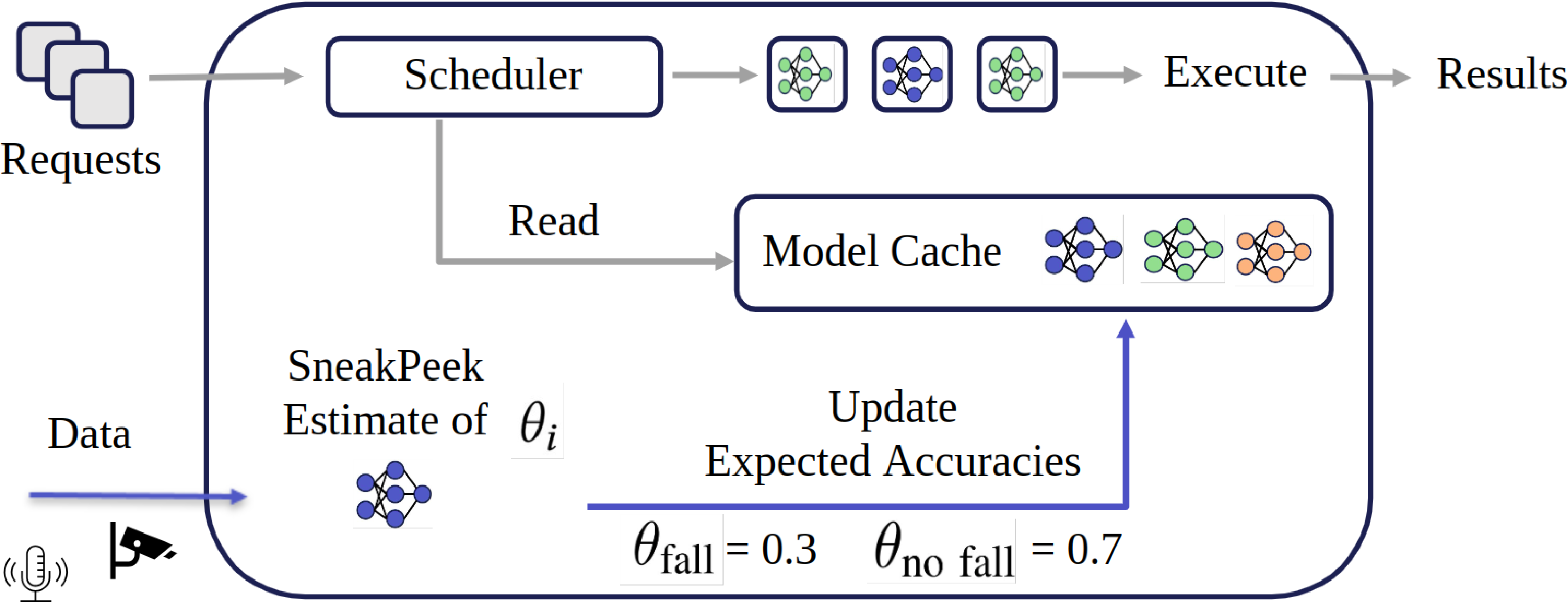}
    \caption{SneakPeek model updating for fall detection}
    \label{fig:sneakpeek-system}
\end{figure}

\subsection{SneakPeek Estimation Scheme}
We require a mechanism for estimating the accuracy of each model, given the data it will operate on.
Our task is to dynamically estimate $\bm{\theta} = \langle\theta_1, ... , \theta_{\mid c \mid}\rangle$ based on the data, which will produce a better estimate of the accuracy of a given model. \\

\noindent
\textbf{Definition 4.1.1.} A \textit{SneakPeek model} is a model that uses real-time data to estimate $\bm{\theta}$. \\

\noindent
SneakPeek models use the underlying data to give us better estimates of model accuracy, which can be used to improve scheduling decisions.
By using SneakPeek models to estimate $\bm{\theta}$ directly, we can perform a single inference and use the result to update model accuracies for every model in $\mathcal{M}_{a_i}$.
At first glance, it appears we can simply assign $\bm{\theta}$ the resulting class probabilities from a SneakPeek model.
However, most machine learning models do not generate true probabilities, in the sense that they do not represent long-run frequencies.
They generate \textit{scores}, then a decision rule is applied to generate a prediction. 
To overcome this limitation, we consider a Bayesian approach for obtaining a (posterior) estimate of $\bm{\theta}$ using the Dirichlet-Multinomial model. \\ 

\noindent
\textbf{Definition 4.1.2.} \textit{SneakPeek probabilities} are the posterior estimates $\bm{\theta} \mid \bm{y}$ computed after using a SneakPeek model to collect the evidence $\bm{y}$. \\

\noindent
We now explicitly define our mechanism for obtaining SneakPeek probabilities, given a model. 
To obtain our posterior estimates of $\bm{\theta}$ we need to define our \textit{prior distribution} and a mechanism for collecting \textit{evidence}. To illustrate this process, we consider a fall detection application in a healthcare setting, where the candidate models must infer whether a subject has fallen over and requires immediate attention. We assume this is a binary classification task (i.e. $|c| = 2$). Figure \ref{fig:sneakpeek-system} shows the inference process at a high-level. 

\noindent
\textbf{Prior Distribution.} Our prior distribution has the form:
\begin{equation}
    \bm{\theta} \sim \textrm{Dirichlet}(\alpha_1, ... , \alpha_{\mid c \mid})
\end{equation}

\noindent
which has hyperparameters $\alpha_i$.
This prior assigns a weight to each class, which should be specified directly by the application owner based on expert experience, rather than taken implicitly from a test set.
For fall detection, a true positive (a subject has fallen) is a relatively rare event, so we may prefer to encode that in our prior distribution. We evaluate different priors in section \ref{sec:prior}. 

\noindent
\textbf{Evidence.} 
We require a low-latency mechanism for generating a multinomial observation for the dirichlet/multinomial model using SneakPeek models.
For the fall detection application, we want to use the data (e.g. maybe a short video clip) to determine which classes are likely present in the data.
We use $k$-nearest neighbors with the original training data to provide this evidence.
For a given data point, we find the $k$ nearest points in the training data and allow each neighboring point to count in our multinomial realization.
For example, if we have $k=5$ and two neighbors have the "no fall" label and three have the "fall" label, then our multinomial evidence would be $\bm{y} = \langle 2, 3 \rangle$.
Note that $k$-nearest-neighbors may not be the fastest algorithm, so approximate nearest neighbors (e.g. locality sensitive hashing) may be required in practice.
An alternative option would be to use a single model and decision rule to produce a unit vector with a single non-zero entry, e.g. $\langle 0, 1 \rangle$ if we believe a fall occurred.
However, this is a low-information update and has the potential to introduce additional error if the predicted class is wrong. 

\noindent
\textbf{Posterior Estimates.} In our model, the Dirichlet distribution is a conjugate prior, so the posterior distribution of $\bm{\theta}$ is:
\begin{equation}
    \bm{\theta} \mid \bm{y} \sim \textrm{Dirichlet}(\alpha_1 + y_1, ... , \alpha_{\mid c \mid} + y_{\mid c \mid})
\end{equation}

\noindent
Therefore, the two hyperparameters in our system are the choice of prior distribution and the number of neighboring points ($k$) to include in the evidence. The value of $k$ can be optimized via standard supervised learning procedures.

An important insight is that existing scheduling algorithms can directly incorporate our SneakPeek modeling to improve accuracy estimation.
The only change required is to include the per-class recall in model profiles.

\section{Model Selection and Scheduling}
\label{sec:scheduling}

\subsection{Proposed Approach}
We require an efficient mechanism for solving the scheduling problem (eq. \ref{np-hard-opt}). 
Toward that end, we propose splitting this problem into two sub-problems: (1) request ordering and (2) model selection.
We first apply the ordering to our available requests, then select a locally-optimal model to satisfy each inference request in the order specified.

\subsubsection{Priority-based Request Ordering}
\begin{figure}
    \centering
    \begin{subfigure}{.23\textwidth}
      \centering
      \includegraphics[width=0.99\linewidth]{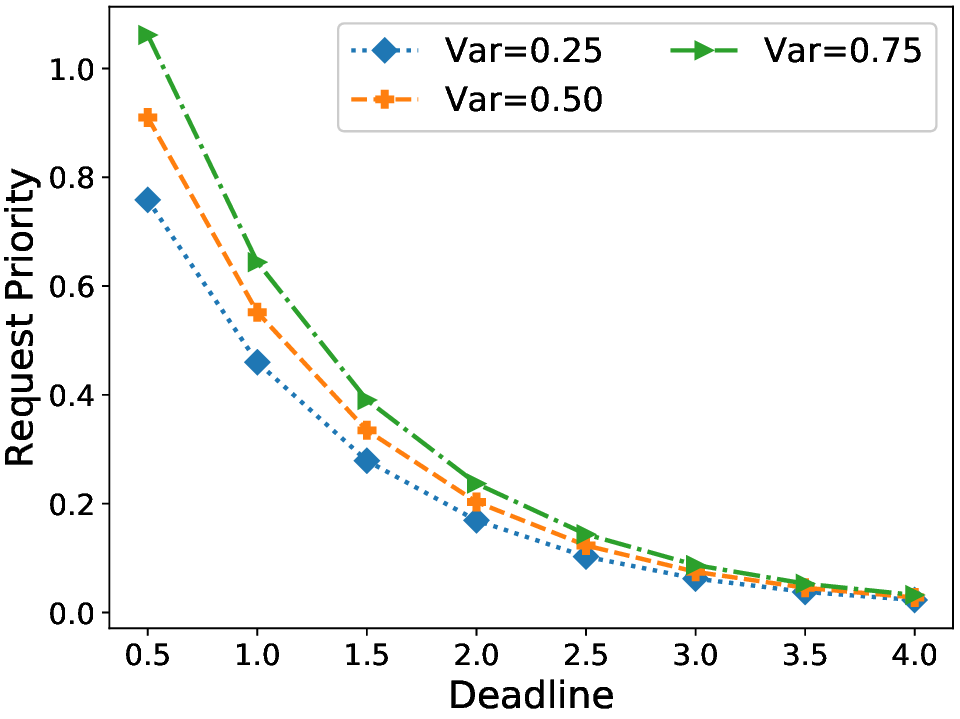}
      \caption{Priority vs. deadline}
      \label{fig:priority-dead}
    \end{subfigure}
    \begin{subfigure}{.23\textwidth}
      \centering
      \includegraphics[width=0.99\linewidth]{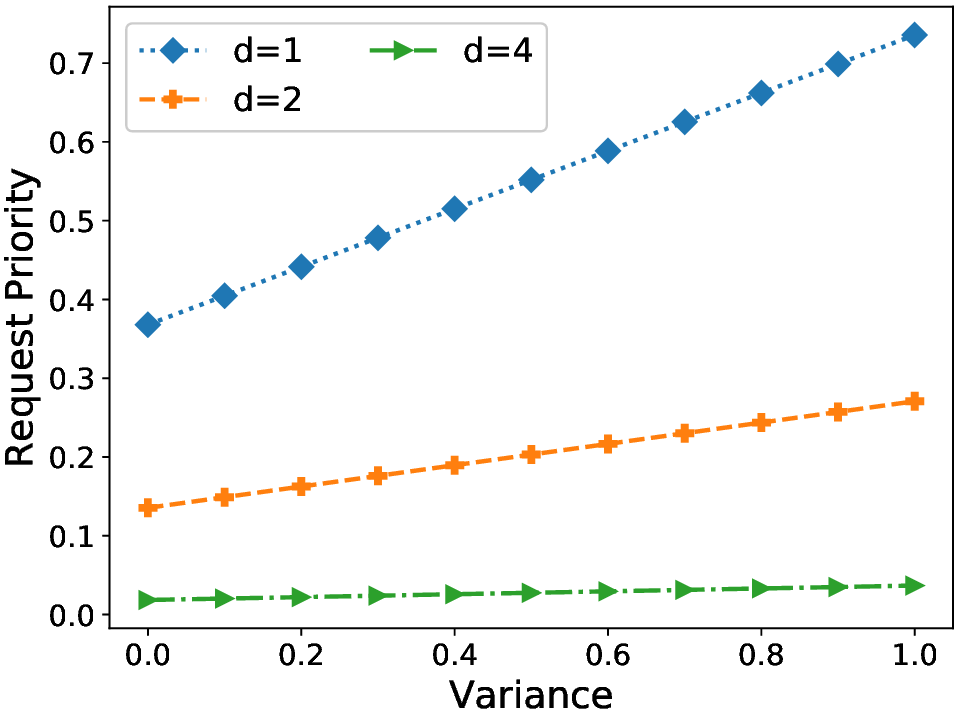}
      \caption{Priority vs. variance}
      \label{fig:priority-var}
    \end{subfigure}
    \caption{Illustration of request priority.}
    \label{fig:priority}
\end{figure}

We require a strategy for ordering the requests prior to model selection.
Common examples from the literature include first come first served (FCFS)~\cite{daisy,layercake} and earliest deadline first (EDF)~\cite{eg-edf}.
We propose a {\em priority-based ordering}, which attempts to dynamically navigate the trade-off between satisfying deadlines and maximizing inference accuracy.
Our goal is to identify a priority function that gives higher values to requests that either (1) have tight deadlines or (2) have a high degree of variability across their choice of models (and thus, have a higher flexibility in model selection).
We define the priority for a given request $r_i$ to be:
\begin{equation}
    \textrm{Priority}(r_i) = \left(1 + \textrm{Var}[Accuracy(\mathcal{M}_{a_i})] \right) e^{-d_i} \label{eq:priority-r}
\end{equation}

\noindent
where $Accuracy(\mathcal{M}_{a_i}) = \{ Accuracy(m_j) \mid m_j \in \mathcal{M}_{a_i} \}$ is the set of model accuracies for $r_i$'s application models.
Requests that have deadlines in the near future or have a high degree of variability\footnote{We use the population variance to compute a priority, so $|\mathcal{M}_{a_i}| = 1 \implies \textrm{Var}[Accuracy(\mathcal{M}_{a_i})] = 0$.} in model accuracy are prioritized.
The intuition is that requests with upcoming deadlines should be prioritized with increasing urgency as we get closer to their deadlines.
At the same time, whenever possible (given the deadlines), our scheduler should focus resources on higher model variance requests.
If there is a low variance in model accuracy, then model selection is unlikely to have much effect on the average utility (which we are trying to maximize).
Figure \ref{fig:priority} illustrates how different deadlines and accuracy variances affect model priority.
When requests are very close to their deadline, the priority increases rapidly. When deadlines are far in the future, the variance in model accuracy has a larger effect on priority.

\subsubsection{Locally-Optimal Model Selection and Placement}
Once requests have been ordered, then for each request $r_i$, we can obtain a locally-optimal solution by selecting the model with the highest utility, that is:
\begin{align}
\label{eq:locally-optimal}
\argmaxA_{m_j \in \mathcal{M}_{a_i}} \; \; \; u_{a_i}(m_j, d_i, t_{i})
\end{align}

\noindent
where $t_i$ is the current time (or the current time plus the expected wait time prior to execution).
Finding the best model is linear in the number of available model variants. 

\subsection{Grouped Scheduling}
Dividing the scheduling problem into request ordering and model selection provides a scalable way to schedule requests in practice. However, using locally-optimal decisions per request prevents schedulers from getting a global view of request dependencies and exploiting model reuse when making decisions.
To get an intuition from optimal solutions, we brute forced solutions to the original scheduling problem in small dimensions (eq. \ref{np-hard-opt}). We observed that \textit{the optimal solutions tend to group requests by application and typically assign the same model to all requests in the group}.
These results suggest that inference batching (or request grouping) is a good strategy, which existing works can only exploit opportunistically if requests happen to be assigned the same model~\cite{proteus, clipper, sponge}.
Using this intuition, we propose {\em grouping requests by application} and applying policies and selection strategies at the {\em group level}, rather than the request level.
This also allows us to exploit model locality and avoid the additional latency associated with swapping models in and out of GPU memory.

We begin by defining the set of groups $\mathcal{G} = \{ g_i \mid g_i \subseteq \mathcal{R} \}$ to be a partition of $\mathcal{R}$.
Each group contains the subset of requests in $\mathcal{R}$ that belong to the same application (and therefore have the same candidate model variants).
More precisely, $r_1, r_2 \in g_i \iff \mathcal{M}_{a_1} = \mathcal{M}_{a_2}$.
Next, we define the priority of a group $g \in \mathcal{G}$ to be:
\begin{equation}
    \textrm{Priority}(g) = \frac{1}{|g|} \sum_{r \in g} \textrm{Priority}(r) \label{eq:priority}
\end{equation}

\noindent
that is, the group priority is simply the mean of the priority of each request in the group.
This approach will attempt to greedily exploit inference batching, but it can also reduce the dimension of the original optimization problem.
The full group-level scheduling algorithm is outlined in Algorithm \ref{algo:group}.
We first check to see if the number of groups is small enough to obtain an exact solution.
If not, we fall back to a locally-optimal solution (applied at the group level), which assigns priorities to each group and schedules all the requests within a group together.
All of our proposed request-based approaches can be solved exactly for a very small number of requests; however, the grouping strategy allows many more scenarios where a solution can be brute forced, since it is a function of the number of applications $|A|$ rather than the number of requests $|\mathcal{R}|$, and $|A| << |\mathcal{R}|$ in practice.

\begin{algorithm}
    \small
    \SetAlgoLined
    \KwIn{$\mathcal{R}$, Model sets $\mathcal{M}_{a_i}$, Brute-Force Threshold $\tau$}
    

    \hspace{0.2cm}
    
    $\mathcal{S}$ $\leftarrow$ $\textrm{Schedule initialized to all zeros}$

    $grouped\_requests$ $\leftarrow$ $\textrm{HashMap}<\textrm{app : request list}>$

    \hspace{0.2cm}

    \If { $\textrm{num\_keys}(grouped\_requests) 
    \leq \tau$ } {
        \Return brute\_force\_solution($grouped\_requests$)
    }
    
    \hspace{0.2cm}
    
    $ordered\_groups$ $\leftarrow$ $\textrm{sorted groups by avg priority (eq. \ref{eq:priority})}$

    \For{$g \textbf{ in } ordered\_groups$} {   
        $m_j = \textrm{ solution to eq. \ref{eq:locally-optimal} using avg group utility}$

        $ordered\_requests$ $\leftarrow$ $\textrm{requests in } g \textrm{ sorted by priority}$
        
        $order \leftarrow 1$
        
        \For{$r_i \textbf{ in } ordered\_requests$} { 
        
            $s_{ij} = order$
            
            $order = order + 1$
        }
    }

    \Return $\; \mathcal{S}$

    \caption{Group-Level Scheduling}
    \label{algo:group}
\end{algorithm}

\subsection{Data-Aware Enhancements}
\begin{figure*}
    \centering
    \includegraphics[width=0.8\linewidth]{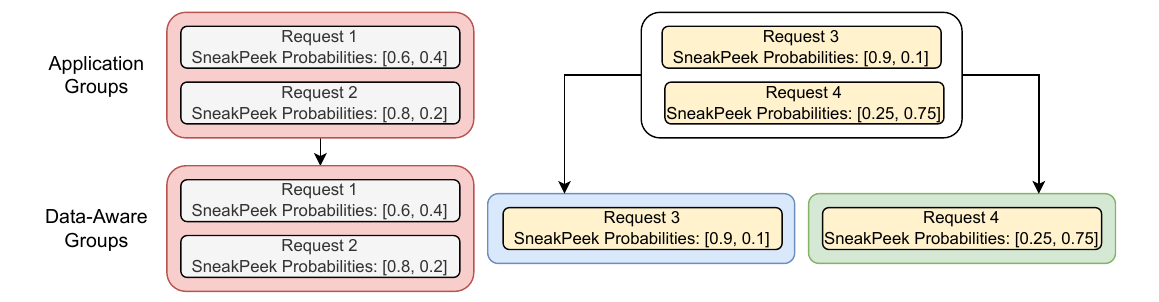}
    \caption{Groups are split based on the computed SneakPeek probabilities}
    \label{fig:group-splitting}
\end{figure*}

We now discuss how our scheduling algorithms can be further improved using the results from our SneakPeek process.

\subsubsection{Short-Circuit Inference}
\label{sec:short}
Since SneakPeek models estimate class probabilities directly, they can be used to provide an additional degree of freedom to schedulers.
If request deadlines are close or a SneakPeek model has very high accuracy, we can consider using the output of the SneakPeek model directly to satisfy the inference request.
To make these decisions, we can simply add the SneakPeek model as a candidate for scheduling (with an inference latency of zero) and let the scheduling algorithms determine if it should be used to maximize overall utility.
However, we must rely on profiled accuracy when making scheduling decisions with SneakPeek models.
We don't want to risk compounding errors by letting a SneakPeek model dynamically estimate its own accuracy based on the data.
Existing scheduling algorithms and our proposed approaches can incorporate short-circuit inference directly, without requiring substantial changes. 

\subsubsection{Enhancements for Grouped Scheduling}
We can further improve our grouped scheduling algorithm by leveraging the fact that model accuracy varies, depending on the class label~\cite{daisy}.
The group scheduler puts all requests for a given application into a single group, which then assigns a single model for all requests in the group. We propose using data-awareness to split groups into subgroups based on the output from SneakPeek models. We will create a subgroup for each target label present in the data. We use the output from SneakPeek models to estimate class membership (indicated by $\theta_i > 0.5$). If $\theta_i < 0.5$ for all $i$, we do not split that request into a subgroup.

Figure \ref{fig:group-splitting} shows how the group splitting works for two binary classification tasks.
In the red group on the left, the SneakPeek probabilities indicate the same class, so no splitting is performed. For the white group on the right, the SneakPeek probabilities indicate different classes are present, so the requests are split into two subgroups. In the case of multi-class classification, the probabilities could also be inconclusive (i.e. $\theta_i < 0.5$ for all $i$) which would not result in any splitting.

\section{Empirical Evaluation}
\label{sec:eval}

\subsection{Methodology}
\noindent
\textbf{Testbed.} We conducted our experiments on a system with an Intel i5 CPU, 32 GiB of main memory, and an NVIDIA RTX 3060 graphics card, which has 12 GiB of memory. 

\begin{table*}[htbp]
\footnotesize
\centering
\caption{Applications and Datasets}
\label{tbl:applications}
\begin{tabular}{|p{1.5cm}|p{4.0cm}|p{3.0cm}|p{6.5cm}|}
        \toprule
        Application & Description & Dataset & Models \\
        \midrule
        Fall \; \; \; \; \; \; \; \; \; Detection & Determine if a subject has fallen (2 classes) & MMAct~\cite{mmact} & X3D~\cite{x3d} for video data (small, medium, and large variants), MiniRocket~\cite{minirocket} for time series, and a fusion model which leverages both modalities~\cite{choi2022multistage} \\
        \midrule
        Voice \; \; \; \; \; \; \; \; \; Commands & Detect keywords spoken by staff members (6 classes) & Speech Commands~\cite{speechcommands} & Howl Framework~\cite{tang-etal-2020-howl} (with LSTM and MobileNet) \\
        \midrule
        Heart \; \; \; \; \; Monitoring & Monitor ECG data for abnormal patterns (7 classes) & MIT-BIH Arrhythmia Database~\cite{mit-heart-data} & EcgResNet34, CNN~\cite{cnn-mit} \\
        \bottomrule
\end{tabular}
\end{table*}

\noindent
\textbf{Datasets and Models.} 
Our evaluation uses systems and applications that are encountered in healthcare settings.
We consider three applications: fall detection, speech commands, and heart monitoring (table \ref{tbl:applications}). 
For fall detection, we have five model variants: 3 video models, one time series model (for accelerometer data), and one fusion model which uses both data modalities.
Since falls are relatively rare events, we stream data such that 95\% of the requests contain true negatives and 5\% of the data contains true positives (a fall).
For the voice command application, we have two model variants and we generate data uniformly across the target classes.
The heart monitoring application also supports two models; we assume arrhythmias are somewhat infrequent and generate true negatives 80\% of the time, with the other 20\% distributed uniformly across the different arrhythmia types. 
Profiled model accuracy is averaged over the data in the test set.
For generating a test set, we follow previously established methodology when available. If no methodology is published, we adopt the standard practice of randomly partitioning the available data points into training and testing sets. 

\noindent
\textbf{Baselines and Metrics.} We compare our proposed approach with several baselines. Each approach consists of an ordering policy-- earliest deadline first (EDF) or our Priority ordering (eq. \ref{eq:priority-r})-- and a model selection strategy:
\begin{itemize}[leftmargin=*,noitemsep]
    \item \textit{Max Accuracy}: Selects the highest accuracy model that can satisfy the request
    \item \textit{Locally-Optimal}: A generalization of several existing works which are deadline-aware~\cite{layercake,slo-aware1,daisy} and select the highest accuracy model that satisfies each deadline.
\end{itemize}
We consider the following combinations in our evaluation:
\begin{itemize}[leftmargin=*,noitemsep]
    \item \textit{MaxAcc-EDF}: Max accuracy model selection + EDF ordering.
    \item \textit{LO-EDF}: Locally-optimal model selection + EDF ordering.
    \item \textit{LO-Priority}: Locally-optimal model selection + priority ordering.
    \item \textit{Grouped}: Our group-level scheduling algorithm (Algorithm~\ref{algo:group}) which groups requests based on target application.
    \item \textit{SneakPeek}: Our group-level scheduling algorithm with SneakPeek data-awareness and short-circuit inference.
\end{itemize}
We evaluate these approaches in terms of scheduling utility, accuracy, and deadline violations.
For utility functions, we use the following penalties:
\begin{itemize}[leftmargin=*,noitemsep]
    \item \textit{Step Function}: Requests that complete after the deadline receive a utility of zero, i.e. $\gamma_{a_i}(d_i, e_i) = \mathbbm{1}_{d_i < e_i}$
    \item \textit{Linear}: Penalty that increases linearly, i.e. $\gamma_{a_i}(d_i, e_{i}) = \; \; \; \; \;$ $\mathbbm{1}_{d_i < e_{i}} \; \textrm{max}(1, \frac{e_i - d_i}{d_i})$
    \item \textit{Sigmoid}: Penalty following a sigmoid curve: $\gamma_{g_i}(d_i, e_{i}) = \mathbbm{1}_{d_i < e_{i}} \; \textrm{max}(1, \frac{1}{1 + (\frac{x}{1-x})^{-3}}$) where $x=1 - \frac{2d_i-e_{i}}{d_i}$
\end{itemize}
\noindent
By default, we use 12 requests (4 for each application) arriving uniformly over a scheduling window of 100ms.
For approximate nearest neighbors, we use the Faiss library~\cite{douze2024faiss} and default to $k=5$. 
Unless otherwise stated, we use an uninformative prior for SneakPeek estimation and default to the sigmoid penalty function.

\subsection{Scheduling Performance}

We first compare our proposed approaches in terms of the resulting utility, accuracy, and deadline violations.
If a deadline is exceeded, the violation time for a given request is the completion time minus the deadline.
For these experiments, we fix the average deadline per request at 150ms. 
To support short-circuit inference, we simply add another model variant for each application, which has a profiled inference latency of zero and the average accuracy of the SneakPeek model.

\begin{figure}
    \centering
        \centering
        \includegraphics[width=0.85\columnwidth]{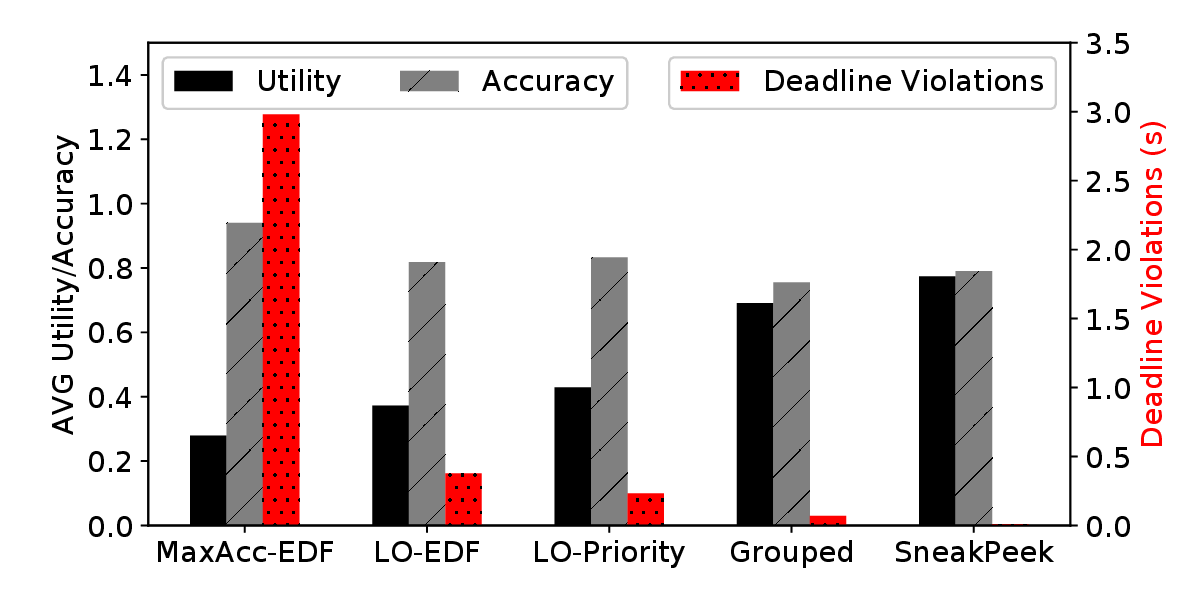}
        \caption{Comparison of schedule utility across approaches}
        \label{fig:main-result}
\end{figure}

Figure \ref{fig:main-result} shows the results for this experiment.
We observe that SneakPeek achieves the highest utility compared to the baselines: a 2x increase in utility compared to LO-EDF.
MaxAcc-EDF always selects the highest accuracy model, which results in both high deadline violations and high average accuracy.
The other approaches have slightly lower accuracy but also substantially lower deadline violations.
The grouped scheduler has the lowest accuracy, but relativly high utility, since it is able to greedily incorporate inference batching to meet more deadlines.
SneakPeek has the fewest deadline violations (almost 0), since it has all the benefits of the grouped scheduler and is able to leverage short-circuit inference to meet deadlines when profiled models cannot.

\subsection{SneakPeek Performance}

\begin{figure}
    \centering
    \includegraphics[width=0.75\linewidth]{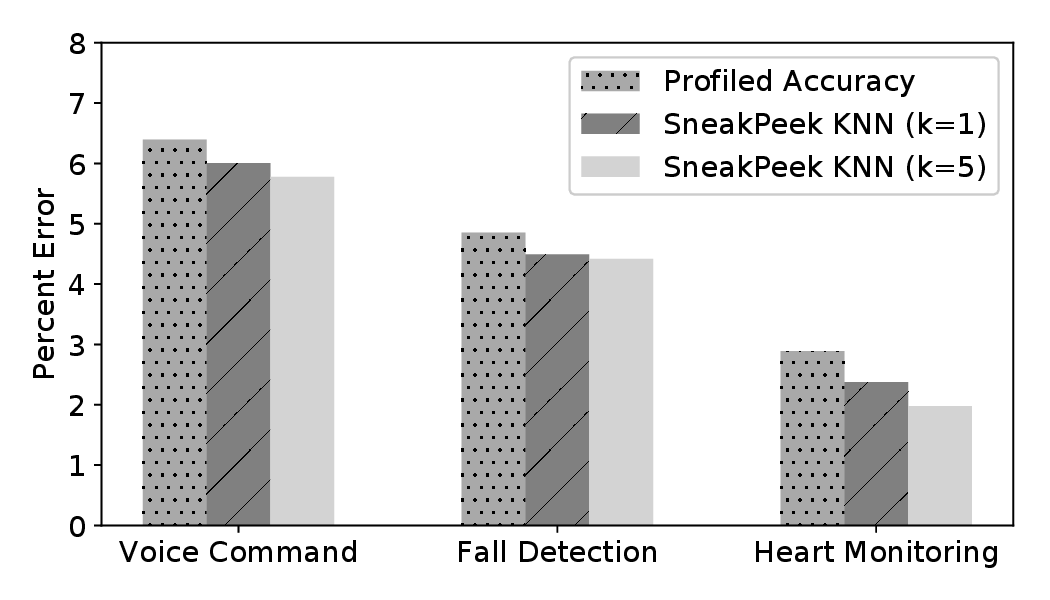}
    \caption{Accuracy estimation error}
    \label{fig:sp-acc}
\end{figure}

\subsubsection{Dynamic Accuracy Estimation}

We now evaluate the degree to which our SneakPeek probabilities reduce error when estimating model accuracy.
We define the "true model accuracy" using the expression in equation \ref{eq:theta-acc}, with $\theta_i = 1$ for the true class label and $\theta_i = 0$ for all other labels.
We use approximate nearest neighbors as our SneakPeek model and evaluate the effect using $k=1$ and $k=5$.
We measure the associated error using all three datasets and restrict ourselves to uninformative priors when generating SneakPeek probabilities.

Figure \ref{fig:sp-acc} shows the results for this experiment.
We observe that in all cases, the SneakPeek probabilities are able to improve the estimates of model accuracy, with $k=5$ performing the best, followed by $k=1$ for approximate nearest neighbors.
Improvements to these accuracy estimates is the main mechanism which allows the data-aware schedulers to increase average schedule utility.

\begin{figure}
    \centering
        \centering
        \includegraphics[width=0.85\columnwidth]{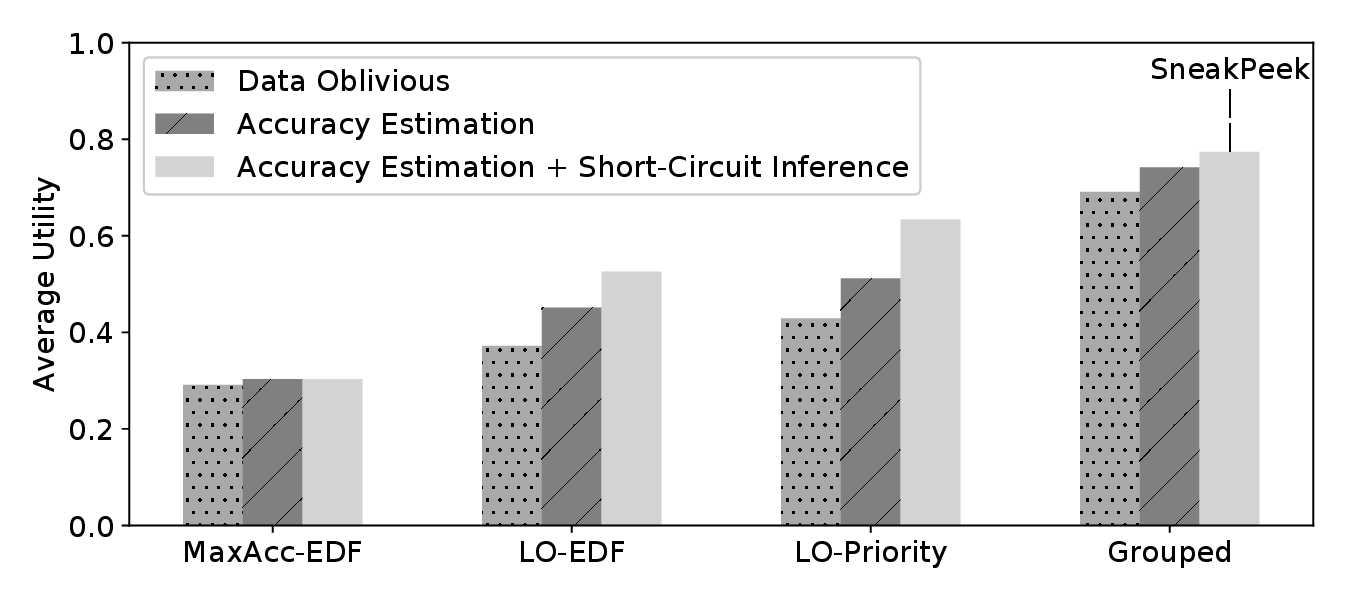}
        \caption{Comparison of schedule utility across approaches}
        \label{fig:data-aware-approaches}
\end{figure}

We also evaluate how the various data-awareness mechanisms incrementally improve the existing approaches (figure \ref{fig:data-aware-approaches}).
First, we observe that the data-oblivious grouped scheduler attains a higher utility than any of the data-aware baselines, showing the benefits of grouping.
In all cases, we observe that incorporating data-awareness in the scheduling process improves schedule utility, showing the benefit of using SneakPeek with any scheduling policy.
We also observe that in most cases, including SneakPeek models as a scheduling option for short-circuit inference improves average schedule utility, although the degree varies across application.
Since most of our approaches involve locally-optimal decisions, the SneakPeek models improve the utility of requests that have been deemed low-priority.
If we cannot satisfy the deadline for these requests, SneakPeek models provide a mechanism for salvaging utility.
These improvements go beyond the benefits of data-awareness, which are provided by using the SneakPeek probabilities to sharpen model accuracy estimates.
Note that MaxAcc-EDF does not benefit from short-circuit inference, since it always selects the model that maximizes accuracy, and SneakPeek is never the most accurate model available.

\begin{figure}
    \centering
    \includegraphics[width=0.6\linewidth]{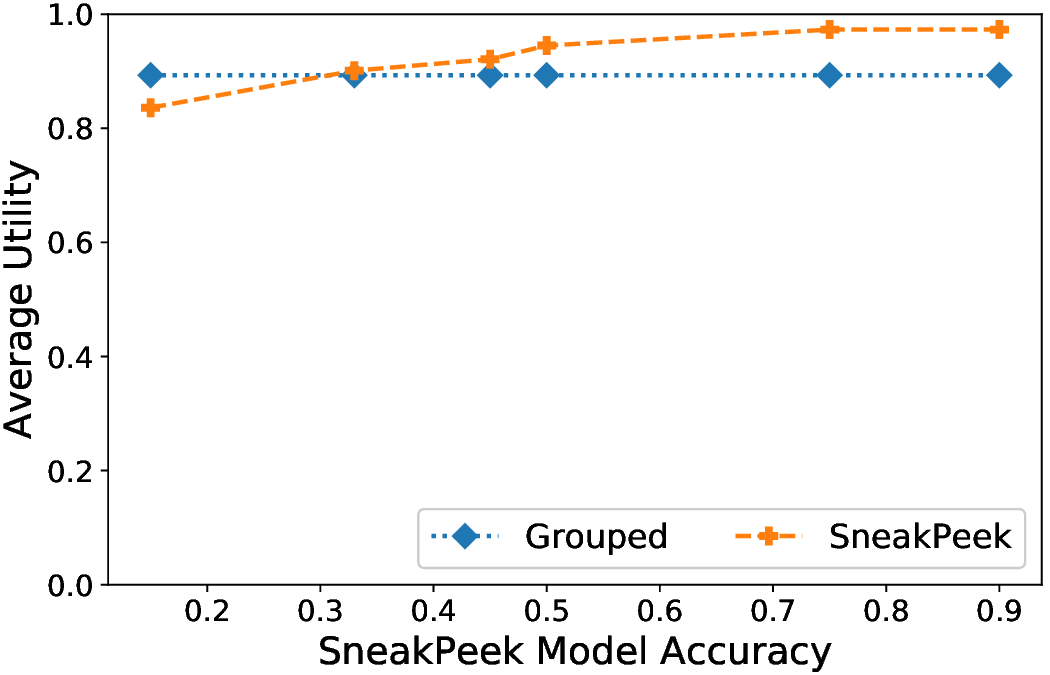}
    \caption{Required accuracy simulation}
    \label{fig:sp-acc-sim}
\end{figure}

\subsubsection{SneakPeek Accuracy Requirements}

Using SneakPeek models to estimate accuracy raises another question: how accurate do SneakPeek models need to be?
In order to answer this question, we create a special SneakPeek model for each application, which randomly returns class probabilities depending on a specified confusion matrix. We simply generate a confusion matrix with the specified accuracy (and uniformly distribute errors across the remaining classes).
Then, given the data point, we randomly generate probabilities using the specified frequencies in the true label row.

Figure \ref{fig:sp-acc-sim} shows that accuracy values above ~30\% provide \textit{some} utility benefits.
For lower accuracy values, we observe that the schedule utility degrades slightly.
These results suggest that SneakPeek models do not require extremely high accuracy, but can begin improving scheduling decisions with moderate inference accuracy.

\subsubsection{Choice of Prior}
\label{sec:prior}

\begin{figure}
    \centering
    \begin{subfigure}{.30\textwidth}
      \centering
      \includegraphics[width=\linewidth]{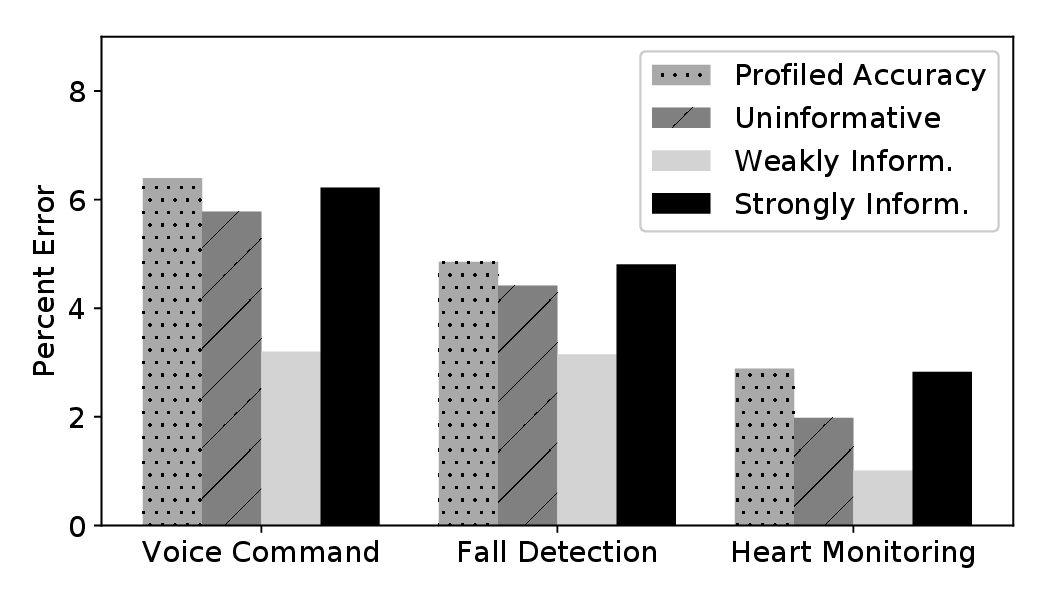}
      \caption{Prior reflects true distribution}
      \label{fig:sp-prior-good}
    \end{subfigure} \vspace{2mm}%
    \begin{subfigure}{.30\textwidth}
      \centering
      \includegraphics[width=\linewidth]{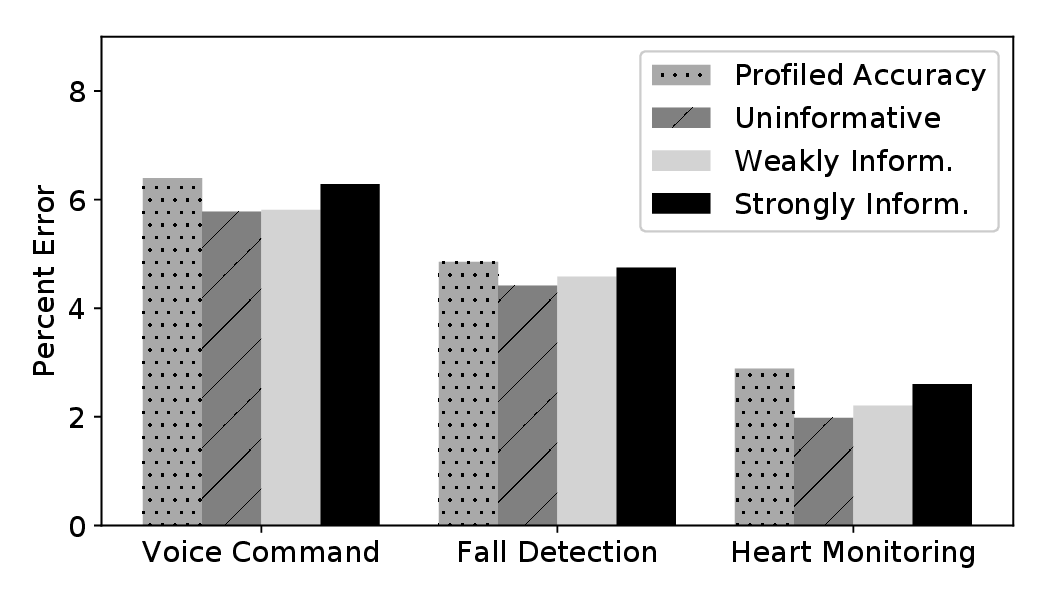}
      \caption{Prior reflects test dataset}
      \label{fig:sp-prior-bad}
    \end{subfigure}
    \caption{Accuracy estimation error with different priors.}
    \label{fig:prior}
\end{figure}

One of the SneakPeek hyperparameters is the choice of prior distribution for $\bm{\theta}$.
We now evaluate our different priors and examine their effects on accuracy estimation. We consider the following prior distributions for SneakPeek estimation: 
\begin{itemize}[leftmargin=*,noitemsep]
    \item \textit{Uninformative}: Does not provide any information regarding the class frequencies. We use the Jeffreys prior, which assigns $\alpha_i = 0.5$ for all $i$. 
    \item \textit{Weakly Informative}: Incorporates expert knowledge, but does not weight it heavily, e.g. by simply assigning $\alpha_i$ the expected frequency of each label.
    \item \textit{Strongly Informative}: Incorporates expert knowledge and weights it heavily, by assigning $\alpha_i$ the expected number of requests with label $i$ in a scheduling window.
\end{itemize}
\noindent
The "true distribution" is defined for each dataset as outlined in the section 5.1.
The distribution of labels in the test set (for profiling) is formed by generating a uniform random sample from the entire dataset for each application. 

Figure \ref{fig:sp-prior-good} shows the resulting effects on accuracy estimation when the prior captures the true distribution in the data.
We observe that our accuracy estimation improves for both uninformative priors and weakly informative priors.
However, the strongly informative priors incur higher error rates, even if the prior matches the true distribution.
This occurs because a strong prior suppresses any signal from the data itself.
By averaging over the heterogeneity in the data, strong priors give us essentially another profiled average. This average may be slightly better than a profiled accuracy with an arbitrary distribution, but it is still suboptimal.

Figure \ref{fig:sp-prior-bad} shows the accuracy when the prior captures the distribution in the test set, rather than the true distribution.
In this case, the uninformative prior provides the lowest error rates. The weakly informative and strongly informative priors are increasingly worse, since they incorporate information that does not represent the true distribution of the data.




\subsection{Sensitivity Analysis}
For these experiments, we present results for LO-EDF (which is representative of existing approaches) and our proposed algorithms, omitting MaxAcc-EDF for clarity.

\subsubsection{Deadlines}

\begin{figure}
    \centering
    \begin{subfigure}{.235\textwidth}
      \centering
      \includegraphics[width=0.99\linewidth]{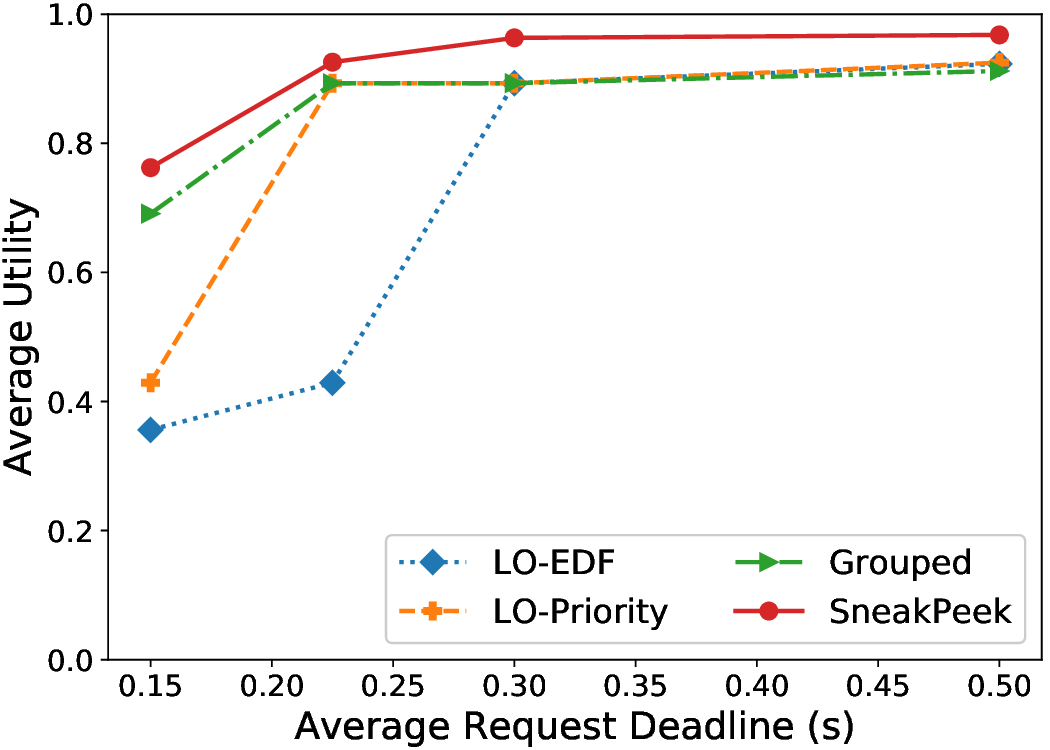}
      \caption{Avg. utility w.r.t. deadlines}
      \label{fig:main-util}
    \end{subfigure} 
    \begin{subfigure}{.235\textwidth}
      \centering
      \includegraphics[width=0.99\linewidth]{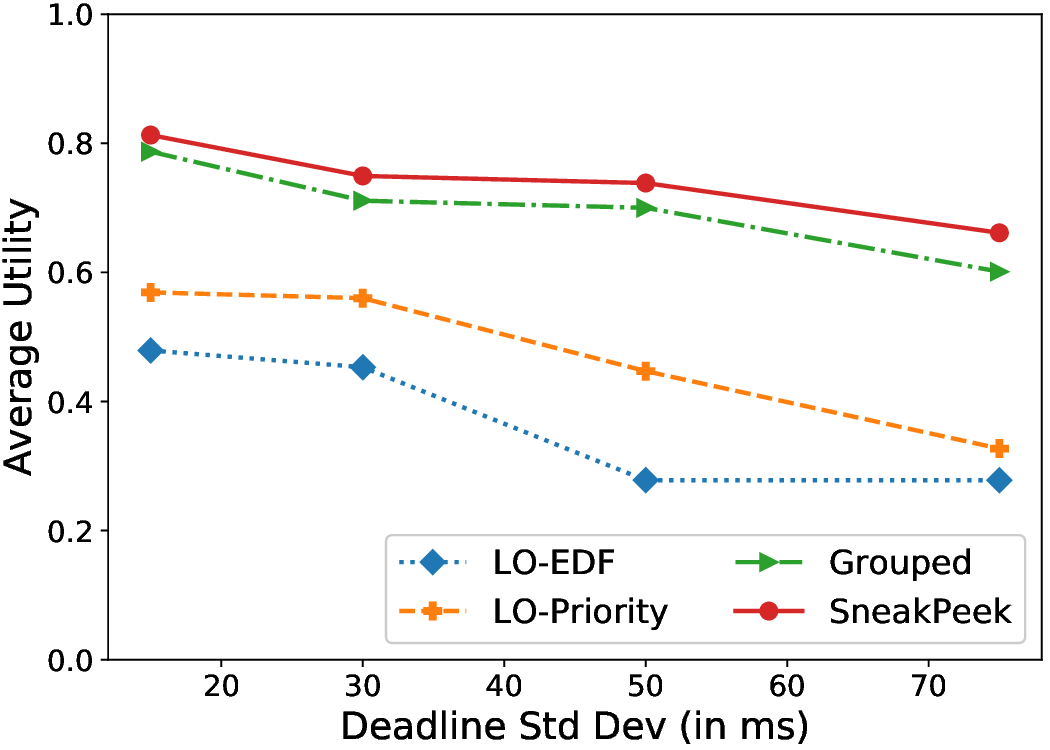}
      \caption{Increasing deadline variance}
      \label{fig:dead-norm}
    \end{subfigure}
    \caption{Impact of deadlines}
    \label{fig:main-dead-r2}
\end{figure}

We now evaluate how different deadline patterns affect the proposed scheduling approaches.

Figure \ref{fig:main-util} examines how utility changes as request deadlines get increasingly larger.
We observe that the grouped scheduler outperforms the proposed baselines for short-term deadlines and the data-aware grouped scheduler consistently outperforms the baselines.
LO-EDF struggles with short deadlines, since EDF does not incorporate information regarding which requests have high accuracy variance. Furthermore, locally-optimal model selection does not exploit inference batching.
After 300 milliseconds, most of the data-oblivious baselines converge to the same utility, since the larger deadlines allow for more flexibility.

We consider another deadline experiment where we increase the {\em variance of the deadlines} across requests.
We generated deadlines following a normal distribution with a mean of 150ms. We then increase the variance of this distribution to observe the effect on utility.
Figure \ref{fig:dead-norm} shows the results for this experiment.
We observe that all approaches slowly degrade as the variance increases.
When there are large differences between request deadlines, the schedulers have fewer degrees of freedom to optimize and must process more requests in the order specified by the deadlines, as the tighter deadline requests become comparatively more urgent.

\subsubsection{Number of Applications}
\begin{figure}
    \centering
    \begin{subfigure}{.235\textwidth}
      \centering
      \includegraphics[width=\linewidth]{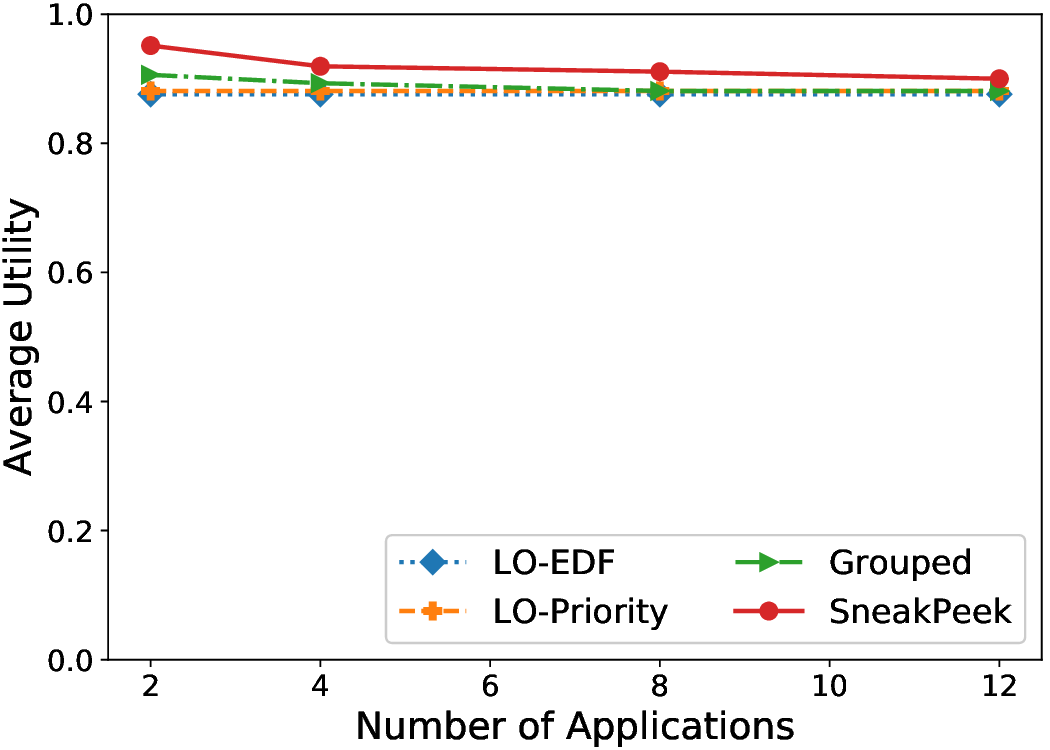}
      \caption{Average Utility}
      \label{fig:apps}
    \end{subfigure} 
    \begin{subfigure}{.235\textwidth}
      \centering
      \includegraphics[width=\linewidth]{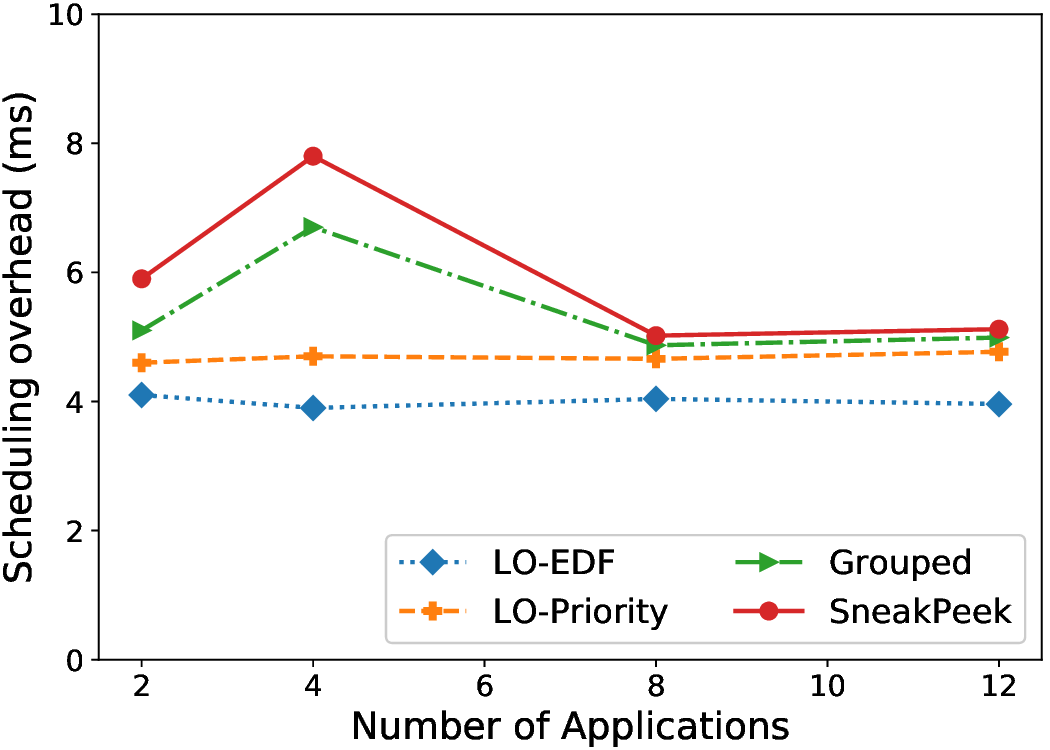}
      \caption{Scheduling Overhead}
      \label{fig:app-scaling}
    \end{subfigure}
    \caption{Effect of number of applications.}
    \label{fig:sensitive-apps}
\end{figure}
We now examine how our approaches behave as the number of applications increases.
Note that increasing the number of applications is not the same as increasing the number of requests.
We fix the number of requests at 24 for the scheduling window and set the average deadline to 200ms (after arrival).
We selected a slightly longer deadline, since the number of requests is higher than in previous experiments.

Figure \ref{fig:apps} shows the results for this experiments.
We observe that the grouped approaches slowly degrade toward the baseline approaches as the number of applications increases (for a fixed number of requests).
This is expected, since increasing the number of applications increases the number of clusters.
So our approach behaves more like LO-Priority as the number of applications increases.
LO-EDF and LO-Priority have consistent performance across an increase in applications, since they make decisions at the request level and the underlying models are the same.

Figure \ref{fig:app-scaling} examines the computational overhead for scheduling as the number of applications increases.  
We first note that all of the approaches are able to produce a schedule in under 10ms.
We observe that for small number of applications, the grouped approach requires additional time, since it is able to solve the assignment problem via brute force.
We also observe an additional increase for the data-aware group-level scheduler, as it creates additional groups based on the target label.
Since the number of requests is fixed, the baseline approaches provide a more consistent overhead.
For larger numbers of applications, the approaches roughly converge.

\subsubsection{Request Arrival Rate}
\begin{figure}
    \centering
    \begin{subfigure}{.235\textwidth}
      \centering
      \includegraphics[width=\linewidth]{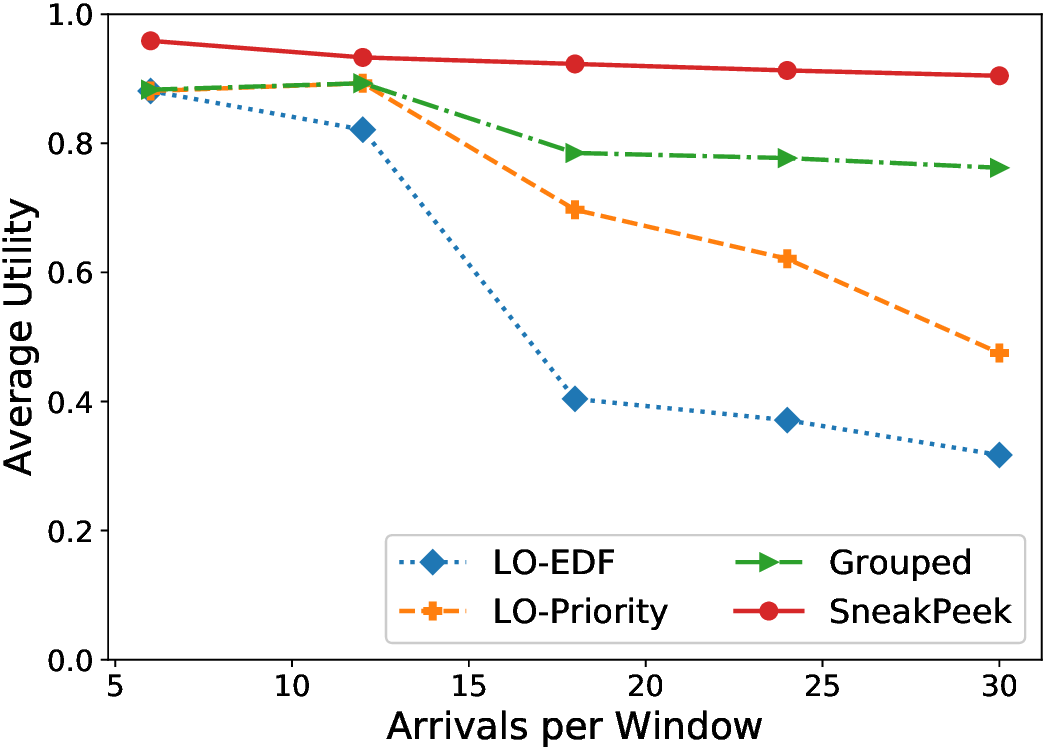}
      \caption{Utility w.r.t. arrival rate}
      \label{fig:arrival}
    \end{subfigure} 
    \begin{subfigure}{.235\textwidth}
      \centering
      \includegraphics[width=\linewidth]{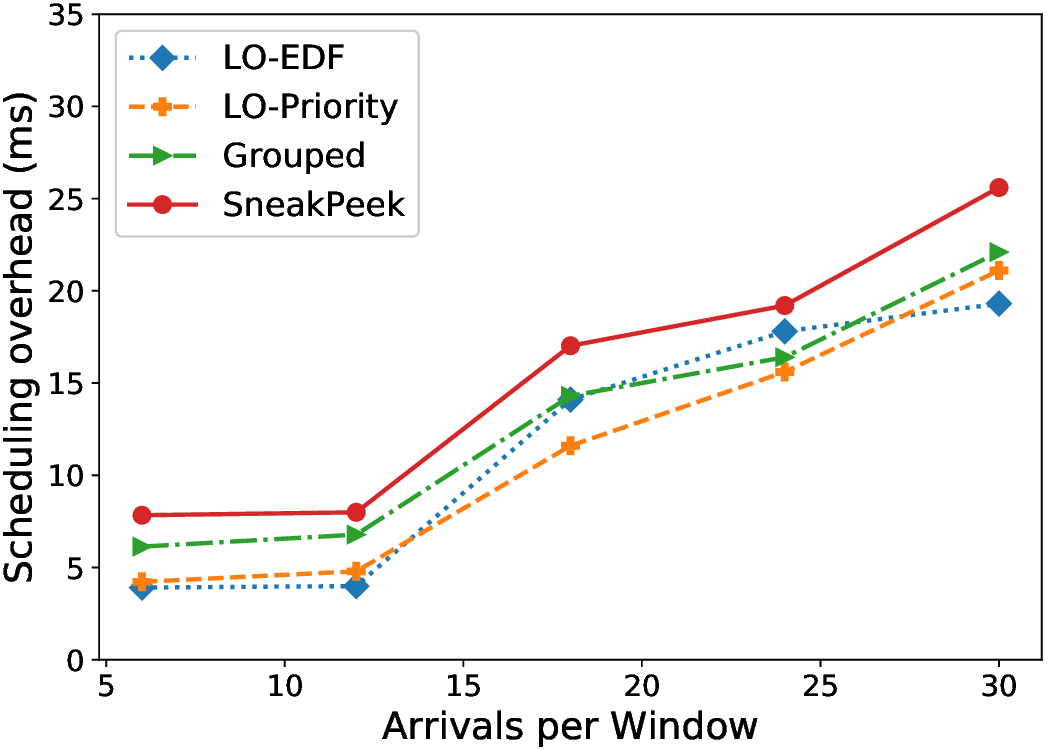}
      \caption{Overhead with Arrival Rate}
      \label{fig:arrival-overhead}
    \end{subfigure}
    \caption{Effect of request arrival rate.}
    \label{fig:sensitive-arrival}
\end{figure}
Our previous experiments assumed an arrival rate of 12 requests, uniformly distributed across a 100ms scheduling window. We now fix the deadline at 200ms and vary the number of requests per window to observe the effects on schedule utility. We can think of this as scaling the number of patients that must be monitored by the existing applications.

Figure \ref{fig:arrival} shows that all approaches slowly experience degraded utility as the number of arrivals increases.
This is expected, since the deadlines are fixed and the amount of inference compute required is increasing.
The schedulers are eventually forced to prioritize requests as the deadlines become tighter.
The grouped schedulers perform the best in this scenario and maintain the highest average utility.
LO-Priority is the next best, since it also prioritizes requests based on the variance in model performance and the anticipated deadline.
LO-EDF is deadline-aware, but does not prioritize requests to account for the increased workload.

Figure \ref{fig:arrival-overhead} examines the computational overhead for scheduling as the number of requests increases. 
For all approaches, we observe an increase in scheduling overhead as the number of requests increases.
This is not surprising, since all of our proposed approaches perform work that is linear in the number of requests.
We also observe a slightly higher overhead for clustering (especially with data-awareness) since the scheduler is able to brute force small numbers of clusters.

\subsubsection{Penalty Function}
\begin{figure}
    \centering
    \begin{subfigure}{.235\textwidth}
      \centering
      \includegraphics[width=\linewidth]{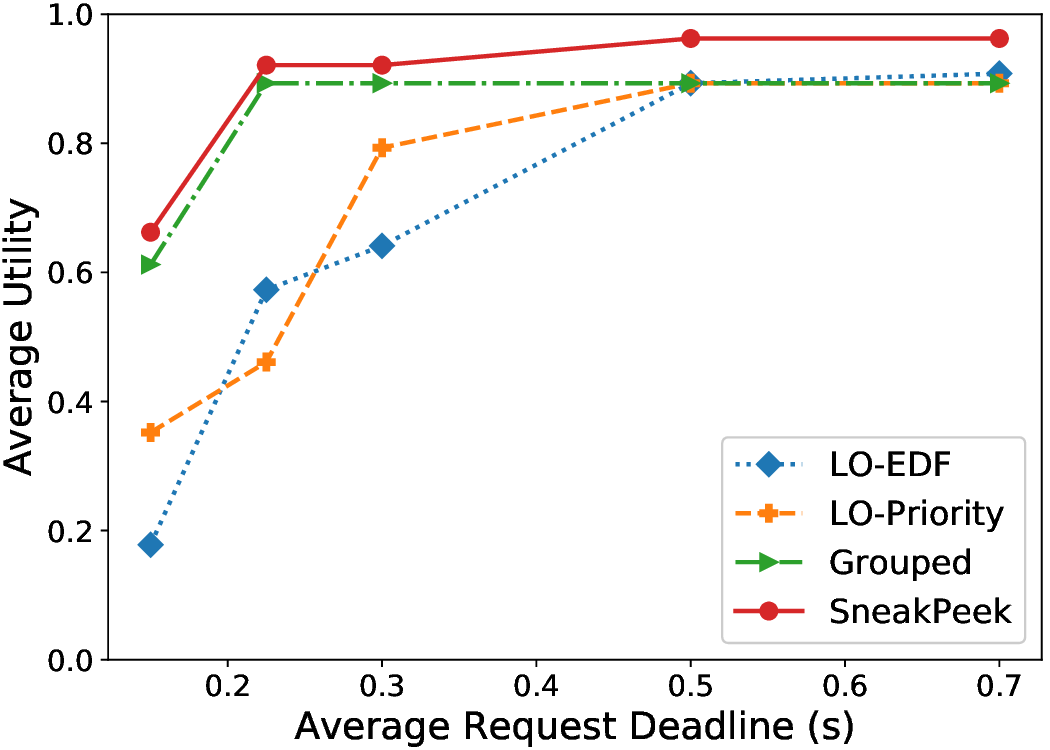}
      \caption{Step Function Penalty}
      \label{fig:penalty-step}
    \end{subfigure} 
    \begin{subfigure}{.235\textwidth}
      \centering
      \includegraphics[width=\linewidth]{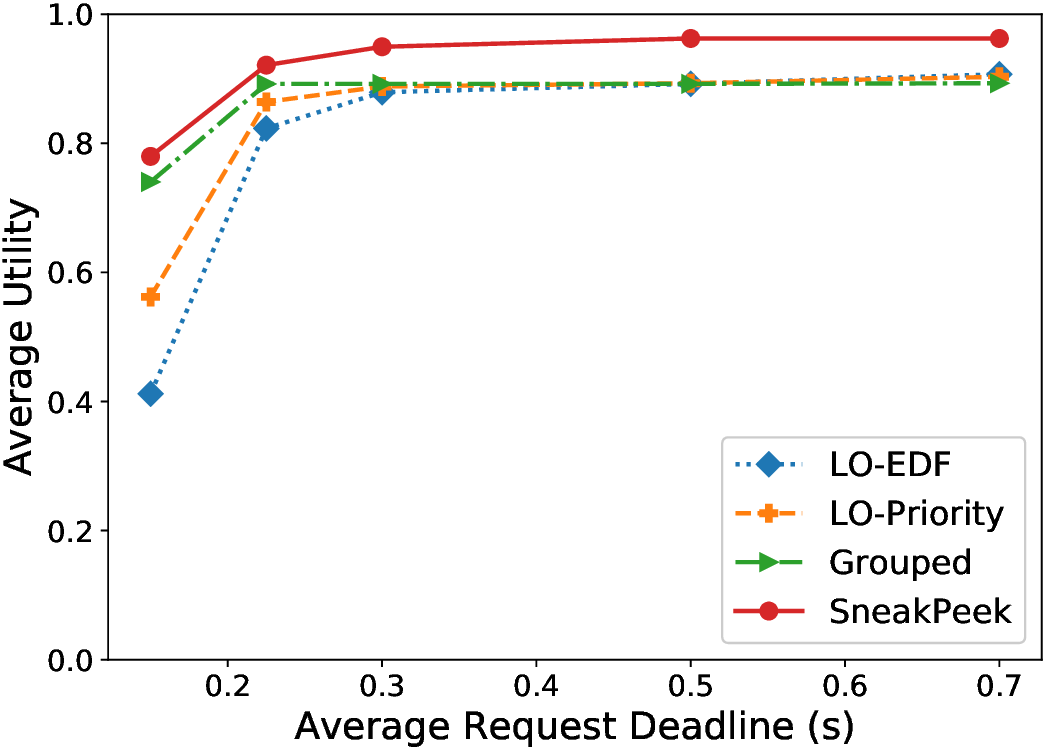}
      \caption{Linear Penalty}
      \label{fig:penalty-lin}
    \end{subfigure}
    \caption{Choice of penalty function.}
    \label{fig:penalty}
\end{figure}
We now examine how various penalty functions affect average scheduling utility.

Figure ~\ref{fig:penalty-step} shows the resulting utilities for a step function.
We observe that our data-aware group scheduler outperforms all the other approaches.
The data oblivious group scheduler also strongly outperforms the baselines when deadlines are short, and is comparable to the baselines when deadlines are longer.
LO-Priority and LO-EDF perform worse in this setting, since they fail to exploit inference batching for early requests. Then the deadlines for later requests are exceeded, which results in a utility of zero. 

Figure ~\ref{fig:penalty-lin} shares a similar pattern with the sigmoid penalty.
SneakPeek consistently outperforms the data-oblivious baselines.
Data-oblivious grouped scheduling also outperforms the baselines when deadlines are short.

\begin{figure}
    \centering
    \includegraphics[width=0.60\columnwidth]{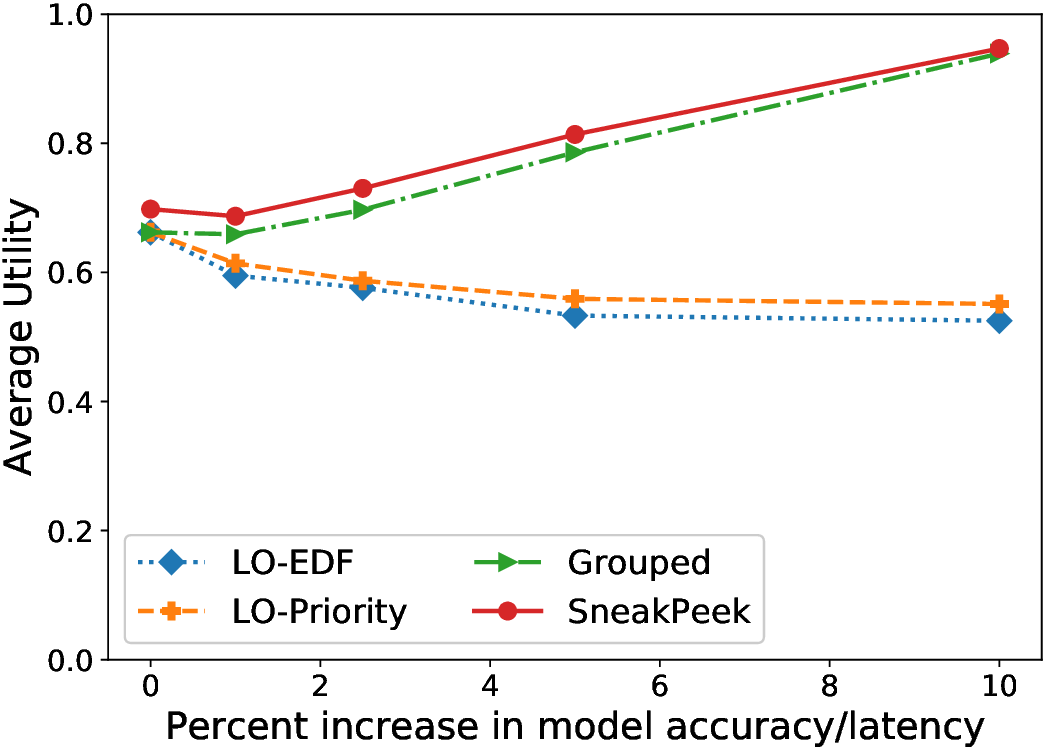}
    \caption{Effect of model heterogeneity.}
    \label{fig:mod-var}
\end{figure}

\subsubsection{Model Variants}
We now explore how different sets of model variants (i.e. choices of $\mathcal{M}_{a_i}$) affect schedule utility.
We exclude the short-circuit approach from this evaluation, as it obscures the direct effects of model heterogeneity when it defers to the SneakPeek model. 
We first created test models that randomly return the correct label following a pre-specified accuracy.
Then, we generate three models for each application: one which has the mean accuracy and inference latency of all models in $\mathcal{M}_a$, and two that we use to increase the variance in model performance for that application. 
At each point, we alter the average accuracy (and latency) by a specified percentage.
For example, if the average model in $\mathcal{M}_{a}$ has 80\% accuracy with an inference latency of 20ms, and we want to increase the variance by 1\%, we will create two models with accuracies of 79.2\% and 80.8\%, and corresponding latencies of 19.8 ms and 20.2 ms.

Figure \ref{fig:mod-var} shows the results for this experiment.
We observe that our proposed grouped approaches benefit substantially as the variance in model performance increases.
The grouped approaches are able to exploit inference batching, which allows for more time to execute high utility models and meet more deadlines.
This suggests that including a diverse set of model variants gives systems the most flexibility, which can help improve utility.
The approaches that do not incorporate grouping obtain lower utilities as the variance increases.
As higher accuracy (and latency) models become available, these locally-optimal approaches select them for the first few requests, but then fail to meet the deadlines for the requests that are later in the ordering.


\section{Multi-Worker Setting}
\label{sec:multi-worker}
\subsection{Problem Definition}
Our problem formulation currently focuses on a system with a single GPU, but our proposed approaches can also be generalized to settings with multiple, heterogeneous workers.
The schedule can be augmented to include a worker in the selection process (indexed by $k$).
For a set of workers $\mathcal{W}$, the (global) optimization objective becomes:
\begin{equation}
    \maxA_{\mathcal{S}} \quad \frac{1}{|\mathcal{R}|} \sum_{i=1}^{|\mathcal{R}|} \sum_{j=1}^{|\mathcal{M}_{a_i}|} \sum_{k=1}^{|\mathcal{W}|} \mathbbm{1}_{s_{ijk} > 0} \; u_{a_i}(m_{j}, d_i, t_{i})
\end{equation}

\noindent
In this setting, each model variant would be profiled on every candidate worker, which would create latency functions that depend on the model and worker.

\subsection{Preliminary Results}
\begin{figure}
    \centering
    \begin{subfigure}{.235\textwidth}
      \centering
      \includegraphics[width=\linewidth]{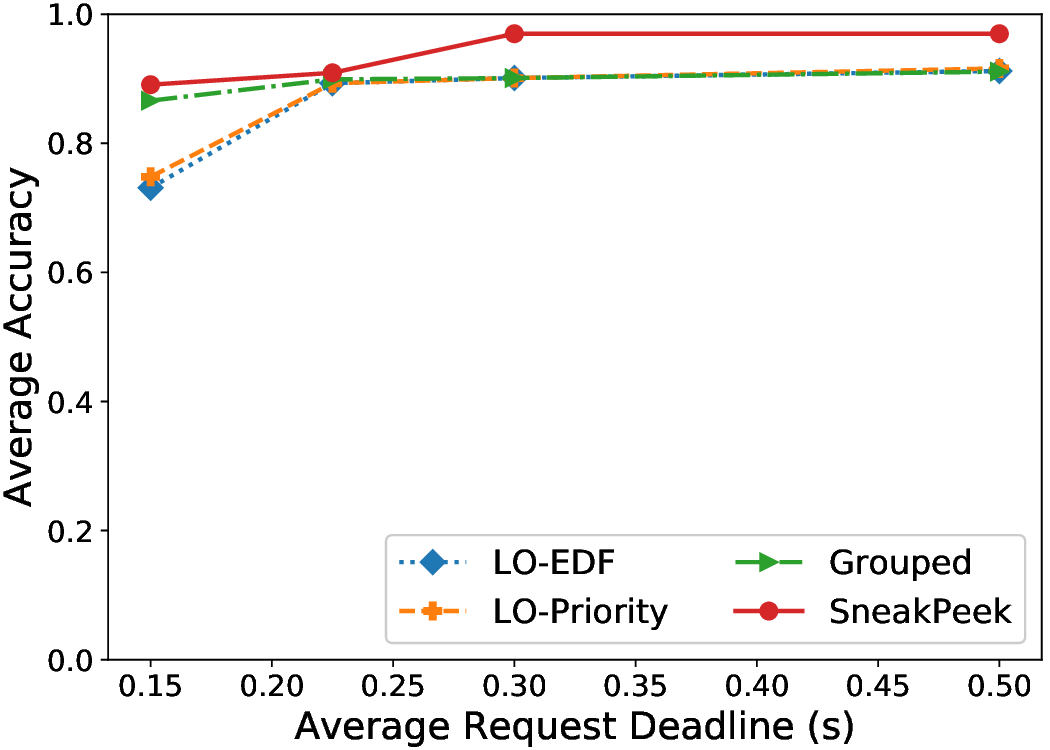}
      \caption{Utility with two workers}
      \label{fig:worker2}
    \end{subfigure} 
    \begin{subfigure}{.235\textwidth}
      \centering
      \includegraphics[width=\linewidth]{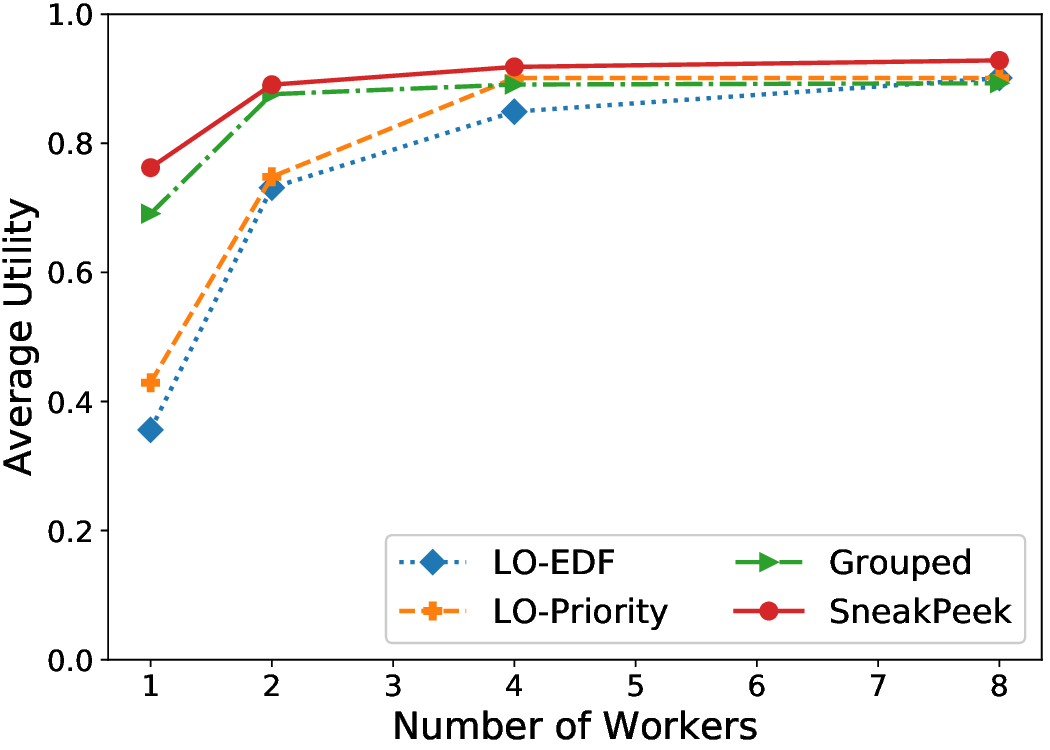}
      \caption{Utility with varying workers}
      \label{fig:worker-vary}
    \end{subfigure}
    \caption{Comparisons with multiple workers.}
    \label{fig:workers}
\end{figure}

We briefly examine how our approach performs in the multi-worker environment by simulating multiple workers on the same hardware used in the single-worker experiments. We ensure that each worker operates independently to minimize resource contention.
In our first experiment (figure \ref{fig:worker2}), we compare schedule utilities with two workers and vary the average deadline. Similar to the single worker case, we observe strong benefits associated with grouped scheduling, which is able to explicitly leverage inference batching. Incorporating data-awareness (SneakPeek) also provides substantial benefit, especially with higher request deadlines.

In figure \ref{fig:worker-vary}, we compare schedule utilities while varying the number of workers. We use an average request deadline of 150 ms. We observe that as the number of workers increases, the relative benefit of our grouped scheduling decreases. This is expected, since the benefit of inference batching decreases if there is less resource contention and fewer requests are assigned to each worker. We also note that the benefit of data-awareness increase slightly with the number of workers. Since deadlines are tight, having more executors allows us to avoid exclusively selecting the fastest model to service the request.

\section{Discussion}
\label{sec:discussion}
Our evaluation highlights the benefits of \textit{data-awareness}, which can be incorporated seamlessly into existing schedulers and allows us to select more accurate models for each request.
We also observed that \textit{grouped} scheduling provides additional benefits by incorporating inference batching decisions into the scheduling process.

Our system performs SneakPeek modeling on an edge system, but a future work could push these operations closer to the data-generating devices.
For example, if this computation can be performed on local IoT devices, data transfers could be avoided, and short-circuit inference could be leveraged when appropriate.
Cloud offloading was excluded from this analysis, since WAN transfers can be costly and applications in healthcare and other domains may be restricted in their ability to offload data.
However, offloading inference to the cloud can also be incorporated into our framework by modifying the latency function $\ell(m_j)$ to include any data transfer time required to run a model in a different location.

Additional modifications may be required to further improve efficiency in the multi-worker setting. For grouped scheduling, we need to enforce a maximum group size and split groups. Having one very large group could lead to load imbalance, so we may require periodic load balancing. Furthermore, data plane optimizations could also be explored to ensure video frames are always available for inference by the time requests are dispatched to workers.

\section{Related Work}

Inference serving (or model serving) has received substantial research attention, with a special focus on cloud deployments~\cite{sponge, eurosys-mss, tolerance-tiers, rl-selection1, compress-sched, mark, infaas, hpdc-cloud}.
The InFaaS system introduced the concept of \textit{model-less} inference serving, where model variants are dynamically selected by the system in addition to serving the request~\cite{infaas}.
Other systems such as MArk attempt to forecast demand and use autoscaling to ensure that performance is consistent across fluctuating workloads~\cite{mark}.
Resource elasticity is a common component in these designs, which is not available in all application settings.
In addition, offloading data to the cloud may be infeasible due to network constraints or privacy requirements.

Several optimizations have been proposed to improve various aspects of the inference serving model (for both edge and cloud deployments).
Several techniques employ \textit{dynamic batching} to reduce inference latency when the scheduler assigns the same model to adjacent requests~\cite{proteus, clipper, sponge}.
However, inference batching can be exploited further when it is incorporated directly in scheduling decisions (e.g. with a grouped scheduler).
A complimentary research direction attempts to improve efficiency by having multiple requests share hardware resources~\cite{interference, 10601534, layercake, hpdc-sched}.
Additional works have considered mechanisms for exploiting multi-modal data sources to improve system efficiency~\cite{daisy, multi-modal1}.
Systems that process multi-modal data often have models with extremely diverse characteristics (accuracy and latency) which can offer additional flexibility to the system~\cite{daisy}.
Efficient interaction with inference serving systems is another important research area, given the overhead of transmitting data and models~\cite{sigmetrics2}.
These optimizations are largely complimentary and can be incorporated directly into our system.

Additional works directly address machine learning inference in hardware-constrained settings, such as the edge~\cite{proteus, slo-aware1, daisy, layercake, loki, sommelier, lao2024hawkvisionlowlatencymodelessedge, edge-adaptor, edge-inference-scheduling,hpdc-edge, octopus}.
A common strategy is to consider accuracy scaling when hardware-scaling is unavailable~\cite{proteus, acc-max1, daisy, layercake, loki,embedded-model-selection}.
The Proteus system maximizing model accuracy, subject to a constraint on the minimum system throughput~\cite{proteus}.
Other edge systems attempt to offload requests when insufficient resources are available locally~\cite{edge-cloud2,layercake}.
The Neurosurgeon system dynamically partitions DNN models to allow parts of the model to execute locally at the edge while the remaining computation is offloaded to the cloud~\cite{neurosurgeon}.
The LayerCake system performs locally-optimal model selection at the edge, while also enumerating candidate models that can be run in the cloud~\cite{layercake}.
Inference pipelines are also commonly deployed at the edge.
Video analytics pipelines such as Chameleon and Ekya leverage spatial-awareness and continuous learning to improve efficiency and accuracy at the edge~\cite{chameleon, ekya}.


\section{Conclusion}

We proposed a scheduling algorithm for hardware-constrained inference serving which implements accuracy scaling and greedily incorporates inference batching into the scheduling process.
We also showed that making decisions based on average model accuracy is suboptimal and proposed SneakPeek: a data-aware approach which attempts to dynamically improve the estimates of model accuracy in real-time.
Our evaluation shows that these techniques obtain higher utility schedules and higher SLO attainment.

\bibliographystyle{IEEEtran}
\bibliography{main-arxiv}

\begin{thebibliography}{10}
\providecommand{\url}[1]{#1}
\csname url@samestyle\endcsname
\providecommand{\newblock}{\relax}
\providecommand{\bibinfo}[2]{#2}
\providecommand{\BIBentrySTDinterwordspacing}{\spaceskip=0pt\relax}
\providecommand{\BIBentryALTinterwordstretchfactor}{4}
\providecommand{\BIBentryALTinterwordspacing}{\spaceskip=\fontdimen2\font plus
\BIBentryALTinterwordstretchfactor\fontdimen3\font minus
  \fontdimen4\font\relax}
\providecommand{\BIBforeignlanguage}[2]{{%
\expandafter\ifx\csname l@#1\endcsname\relax
\typeout{** WARNING: IEEEtran.bst: No hyphenation pattern has been}%
\typeout{** loaded for the language `#1'. Using the pattern for}%
\typeout{** the default language instead.}%
\else
\language=\csname l@#1\endcsname
\fi
#2}}
\providecommand{\BIBdecl}{\relax}
\BIBdecl

\bibitem{8327042}
K.~Hazelwood, S.~Bird, D.~Brooks, S.~Chintala, U.~Diril, D.~Dzhulgakov,
  M.~Fawzy, B.~Jia, Y.~Jia, A.~Kalro, J.~Law, K.~Lee, J.~Lu, P.~Noordhuis,
  M.~Smelyanskiy, L.~Xiong, and X.~Wang, ``Applied machine learning at
  facebook: A datacenter infrastructure perspective,'' in \emph{2018 IEEE
  International Symposium on High Performance Computer Architecture (HPCA)},
  2018, pp. 620--629.

\bibitem{infaas}
F.~Romero, Q.~Li, N.~J. Yadwadkar, and C.~Kozyrakis, ``{INFaaS}: Automated
  model-less inference serving,'' in \emph{2021 USENIX Annual Technical
  Conference (USENIX ATC 21)}.\hskip 1em plus 0.5em minus 0.4em\relax USENIX
  Association, Jul. 2021, pp. 397--411.

\bibitem{model-serving-health}
S.~Hong, Y.~Xu, A.~Khare, S.~Priambada, K.~Maher, A.~Aljiffry, J.~Sun, and
  A.~Tumanov, ``Holmes: Health online model ensemble serving for deep learning
  models in intensive care units,'' in \emph{Proceedings of the 26th ACM SIGKDD
  International Conference on Knowledge Discovery \& Data Mining}, ser. KDD
  '20.\hskip 1em plus 0.5em minus 0.4em\relax New York, NY, USA: Association
  for Computing Machinery, 2020, p. 1614–1624.

\bibitem{9773234}
L.~Ke, U.~Gupta, M.~Hempstead, C.-J. Wu, H.-H.~S. Lee, and X.~Zhang,
  ``Hercules: Heterogeneity-aware inference serving for at-scale personalized
  recommendation,'' in \emph{2022 IEEE International Symposium on
  High-Performance Computer Architecture (HPCA)}, 2022, pp. 141--154.

\bibitem{mobile-serve}
Z.~Fang, D.~Hong, and R.~K. Gupta, ``Serving deep neural networks at the cloud
  edge for vision applications on mobile platforms,'' in \emph{Proceedings of
  the 10th ACM Multimedia Systems Conference}, ser. MMSys '19.\hskip 1em plus
  0.5em minus 0.4em\relax New York, NY, USA: Association for Computing
  Machinery, 2019, p. 36–47.

\bibitem{octopus}
Z.~Zhang, Y.~Zhao, and J.~Liu, ``Octopus: Slo-aware progressive inference
  serving via deep reinforcement learning in multi-tenant edge cluster,'' in
  \emph{Service-Oriented Computing}, F.~Monti, S.~Rinderle-Ma,
  A.~Ruiz~Cort{\'e}s, Z.~Zheng, and M.~Mecella, Eds.\hskip 1em plus 0.5em minus
  0.4em\relax Cham: Springer Nature Switzerland, 2023, pp. 242--258.

\bibitem{daisy}
J.~Wolfrath, A.~Achanta, and A.~Chandra, ``Leveraging multi-modal data for
  efficient edge inference serving,'' in \emph{2024 IEEE/ACM 24th International
  Symposium on Cluster, Cloud and Internet Computing (CCGrid)}, 2024, pp.
  408--417.

\bibitem{sponge}
K.~Razavi, S.~Ghafouri, M.~Mühlhäuser, P.~Jamshidi, and L.~Wang, ``Sponge:
  Inference serving with dynamic slos using in-place vertical scaling,'' in
  \emph{Proceedings of the 4th Workshop on Machine Learning and Systems}, ser.
  EuroSys ’24.\hskip 1em plus 0.5em minus 0.4em\relax ACM, Apr. 2024, p.
  184–191.

\bibitem{eurosys-mss}
D.~Mendoza, F.~Romero, and C.~Trippel, ``Model selection for latency-critical
  inference serving,'' in \emph{Proceedings of the Nineteenth European
  Conference on Computer Systems}, ser. EuroSys '24.\hskip 1em plus 0.5em minus
  0.4em\relax New York, NY, USA: Association for Computing Machinery, 2024, p.
  1016–1038.

\bibitem{tolerance-tiers}
M.~Halpern, B.~Boroujerdian, T.~Mummert, E.~Duesterwald, and V.~J. Reddi, ``One
  size does not fit all: Quantifying and exposing the accuracy-latency
  trade-off in machine learning cloud service apis via tolerance tiers,'' in
  \emph{2019 IEEE International Symposium on Performance Analysis of Systems
  and Software (ISPASS)}, Los Alamitos, CA, USA, 2019.

\bibitem{rl-selection1}
Z.~Fang, T.~Yu, O.~J. Mengshoel, and R.~K. Gupta, ``Qos-aware scheduling of
  heterogeneous servers for inference in deep neural networks,'' in
  \emph{Proceedings of the 2017 ACM on Conference on Information and Knowledge
  Management}, ser. CIKM '17.\hskip 1em plus 0.5em minus 0.4em\relax New York,
  NY, USA: Association for Computing Machinery, 2017, p. 2067–2070.

\bibitem{compress-sched}
T.~D.~S. Barros \emph{et~al.}, ``Scheduling with fully compressible tasks:
  Application to deep learning inference with neural network compression,'' in
  \emph{2024 IEEE/ACM 24th International Symposium on Cluster, Cloud and
  Internet Computing (CCGrid)}, 2024, pp. 327--336.

\bibitem{mark}
C.~Zhang, M.~Yu, W.~Wang, and F.~Yan, ``{MArk}: Exploiting cloud services for
  {Cost-Effective}, {SLO-Aware} machine learning inference serving,'' in
  \emph{2019 USENIX Annual Technical Conference (USENIX ATC 19)}.\hskip 1em
  plus 0.5em minus 0.4em\relax Renton, WA: USENIX Association, Jul. 2019, pp.
  1049--1062.

\bibitem{proteus}
S.~Ahmad, H.~Guan, B.~D. Friedman, T.~Williams, R.~K. Sitaraman, and T.~Woo,
  ``Proteus: A high-throughput inference-serving system with accuracy
  scaling,'' in \emph{Proceedings of the 29th ACM International Conference on
  Architectural Support for Programming Languages and Operating Systems, Volume
  1}, ser. ASPLOS '24.\hskip 1em plus 0.5em minus 0.4em\relax New York, NY,
  USA: Association for Computing Machinery, 2024, p. 318–334.

\bibitem{loki}
S.~Ahmad, H.~Guan, and R.~K. Sitaraman, ``Loki: A system for serving ml
  inference pipelines with hardware and accuracy scaling,'' in
  \emph{Proceedings of the 33rd International Symposium on High-Performance
  Parallel and Distributed Computing}, ser. HPDC '24.\hskip 1em plus 0.5em
  minus 0.4em\relax New York, NY, USA: Association for Computing Machinery,
  2024, p. 267–280.

\bibitem{jellyfish}
V.~Nigade, P.~Bauszat, H.~Bal, and L.~Wang, ``Jellyfish: Timely inference
  serving for dynamic edge networks,'' in \emph{2022 IEEE Real-Time Systems
  Symposium (RTSS)}, 2022, pp. 277--290.

\bibitem{killer-app}
G.~Ananthanarayanan \emph{et~al.}, ``Real-time video analytics: The killer app
  for edge computing,'' \emph{Computer}, 2017.

\bibitem{Olatunji2019}
I.~E. Olatunji and C.-H. Cheng, \emph{Video Analytics for Visual Surveillance
  and Applications: An Overview and Survey}.\hskip 1em plus 0.5em minus
  0.4em\relax Springer, 2019.

\bibitem{ieee-health-survey}
S.~Cristina, V.~Despotovic, R.~Pérez-Rodríguez, and S.~Aleksic, ``Audio- and
  video-based human activity recognition systems in healthcare,'' \emph{IEEE
  Access}, vol.~12, pp. 8230--8245, 2024.

\bibitem{Hassan2019}
M.~K. Hassan, A.~I. El~Desouky, S.~M. Elghamrawy, and A.~M. Sarhan, \emph{Big
  Data Challenges and Opportunities in Healthcare Informatics and Smart
  Hospitals}.\hskip 1em plus 0.5em minus 0.4em\relax Cham: Springer
  International Publishing, 2019.

\bibitem{ad1}
S.~C. Bollepalli, R.~K. Sevakula, W.~M. Au-Yeung, M.~B. Kassab, F.~M. Merchant,
  G.~Bazoukis, R.~Boyer, E.~M. Isselbacher, and A.~A. Armoundas,
  ``{{R}eal-{T}ime {A}rrhythmia {D}etection {U}sing {H}ybrid {C}onvolutional
  {N}eural {N}etworks},'' \emph{J Am Heart Assoc}, vol.~10, no.~23, p. e023222,
  Dec 2021.

\bibitem{seizure1}
K.~Rasheed \emph{et~al.}, ``Machine learning for predicting epileptic seizures
  using eeg signals: A review,'' \emph{IEEE Reviews in Biomedical Engineering},
  vol.~14, pp. 139--155, 2021.

\bibitem{seizure2}
M.~K. Siddiqui \emph{et~al.}, ``A review of epileptic seizure detection using
  machine learning classifiers,'' \emph{Brain Informatics}, vol.~7, no.~1,
  p.~5, May 2020.

\bibitem{resp1}
H.~Burdick, C.~Lam, S.~Mataraso, A.~Siefkas, G.~Braden, R.~P. Dellinger,
  A.~McCoy, J.-L. Vincent, A.~Green-Saxena, G.~Barnes, J.~Hoffman, J.~Calvert,
  E.~Pellegrini, and R.~Das, ``Prediction of respiratory decompensation in
  covid-19 patients using machine learning: The ready trial,'' \emph{Computers
  in Biology and Medicine}, vol. 124, p. 103949, 2020.

\bibitem{resp2}
B.~Friedman, D.~Fuckert, M.~Jahrsdoerfer, R.~Magness, E.~S. Patterson, R.~Syed,
  and J.~R. Zaleski, ``Identifying and monitoring respiratory compromise:
  Report from the rules and algorithms working group,'' \emph{Biomedical
  Instrumentation \&amp; Technology}, vol.~53, no.~2, pp. 110--123, 2019.

\bibitem{springer-scheduling}
B.~Chen, C.~N. Potts, and G.~J. Woeginger, \emph{A Review of Machine
  Scheduling: Complexity, Algorithms and Approximability}.\hskip 1em plus 0.5em
  minus 0.4em\relax Boston, MA: Springer US, 1998, pp. 1493--1641.

\bibitem{siam-scheduling}
J.~Bruno and P.~Downey, ``Complexity of task sequencing with deadlines, set-up
  times and changeover costs,'' \emph{SIAM Journal on Computing}, vol.~7,
  no.~4, pp. 393--404, 1978.

\bibitem{acc-def}
\BIBentryALTinterwordspacing
M.~Grandini, E.~Bagli, and G.~Visani, ``Metrics for multi-class classification:
  an overview,'' 2020. [Online]. Available:
  \url{https://arxiv.org/abs/2008.05756}
\BIBentrySTDinterwordspacing

\bibitem{layercake}
S.~Ogden and T.~Guo, ``Layercake: Efficient inference serving with cloud and
  mobile resources,'' in \emph{2023 23nd IEEE International Symposium on
  Cluster, Cloud and Internet Computing (CCGrid)}, 2023.

\bibitem{eg-edf}
D.~Mendoza, F.~Romero, and C.~Trippel, ``Model selection for latency-critical
  inference serving,'' in \emph{Proceedings of the Nineteenth European
  Conference on Computer Systems}, ser. EuroSys '24.\hskip 1em plus 0.5em minus
  0.4em\relax New York, NY, USA: Association for Computing Machinery, 2024, p.
  1016–1038.

\bibitem{clipper}
D.~Crankshaw, X.~Wang, G.~Zhou, M.~J. Franklin, J.~E. Gonzalez, and I.~Stoica,
  ``Clipper: A {Low-Latency} online prediction serving system,'' in \emph{14th
  USENIX Symposium on Networked Systems Design and Implementation (NSDI
  17)}.\hskip 1em plus 0.5em minus 0.4em\relax Boston, MA: USENIX Association,
  Mar. 2017, pp. 613--627.

\bibitem{mmact}
Q.~Kong, Z.~Wu, Z.~Deng, M.~Klinkigt, B.~Tong, and T.~Murakami, ``Mmact: A
  large-scale dataset for cross modal human action understanding,'' in
  \emph{Proceedings of the IEEE/CVF International Conference on Computer Vision
  (ICCV)}, October 2019.

\bibitem{x3d}
C.~Feichtenhofer, ``X3d: Expanding architectures for efficient video
  recognition,'' in \emph{2020 IEEE/CVF Conference on Computer Vision and
  Pattern Recognition (CVPR)}.\hskip 1em plus 0.5em minus 0.4em\relax Los
  Alamitos, CA, USA: IEEE Computer Society, jun 2020, pp. 200--210.

\bibitem{minirocket}
A.~Dempster, D.~F. Schmidt, and G.~I. Webb, ``Minirocket: A very fast (almost)
  deterministic transform for time series classification,'' ser. KDD '21.\hskip
  1em plus 0.5em minus 0.4em\relax New York, NY, USA: Association for Computing
  Machinery, 2021, p. 248–257.

\bibitem{choi2022multistage}
H.~Choi, A.~Beedu, H.~Haresamudram, and I.~Essa, ``Multi-stage based feature
  fusion of multi-modal data for human activity recognition,'' 2022.

\bibitem{speechcommands}
P.~Warden, ``Speech commands: A public dataset for single-word speech
  recognition.'' 2017.

\bibitem{tang-etal-2020-howl}
R.~Tang, J.~Lee, A.~Razi, J.~Cambre, I.~Bicking, J.~Kaye, and J.~Lin, ``Howl: A
  deployed, open-source wake word detection system,'' in \emph{Proceedings of
  Second Workshop for NLP Open Source Software (NLP-OSS)}.\hskip 1em plus 0.5em
  minus 0.4em\relax Association for Computational Linguistics, Nov. 2020, pp.
  61--65.

\bibitem{mit-heart-data}
A.~L. Goldberger, L.~A. Amaral, L.~Glass, J.~M. Hausdorff, P.~C. Ivanov, R.~G.
  Mark, J.~E. Mietus, G.~B. Moody, C.~K. Peng, and H.~E. Stanley,
  ``\BIBforeignlanguage{en}{{PhysioBank}, {PhysioToolkit}, and {PhysioNet}:
  components of a new research resource for complex physiologic signals},''
  \emph{\BIBforeignlanguage{en}{Circulation}}, vol. 101, no.~23, pp. E215--20,
  Jun. 2000.

\bibitem{cnn-mit}
\BIBentryALTinterwordspacing
T.~J. Jun, H.~M. Nguyen, D.~Kang, D.~Kim, D.~Kim, and Y.~Kim, ``{ECG}
  arrhythmia classification using a 2-d convolutional neural network,''
  \emph{CoRR}, vol. abs/1804.06812, 2018. [Online]. Available:
  \url{http://arxiv.org/abs/1804.06812}
\BIBentrySTDinterwordspacing

\bibitem{slo-aware1}
W.~Seo, S.~Cha, Y.~Kim, J.~Huh, and J.~Park, ``Slo-aware inference scheduler
  for heterogeneous processors in edge platforms,'' \emph{ACM Trans. Archit.
  Code Optim.}, vol.~18, no.~4, jul 2021.

\bibitem{douze2024faiss}
M.~Douze, A.~Guzhva, C.~Deng, J.~Johnson, G.~Szilvasy, P.-E. Mazaré,
  M.~Lomeli, L.~Hosseini, and H.~Jégou, ``The faiss library,'' 2024.

\bibitem{hpdc-cloud}
B.~Li, S.~Samsi, V.~Gadepally, and D.~Tiwari, ``Kairos: Building cost-efficient
  machine learning inference systems with heterogeneous cloud resources,'' in
  \emph{Proceedings of the 32nd International Symposium on High-Performance
  Parallel and Distributed Computing}, ser. HPDC '23.\hskip 1em plus 0.5em
  minus 0.4em\relax Association for Computing Machinery, 2023, p. 3–16.

\bibitem{interference}
D.~Mendoza, F.~Romero, Q.~Li, N.~J. Yadwadkar, and C.~Kozyrakis,
  ``Interference-aware scheduling for inference serving,'' in \emph{Proceedings
  of the 1st Workshop on Machine Learning and Systems}, ser. EuroMLSys
  '21.\hskip 1em plus 0.5em minus 0.4em\relax New York, NY, USA: Association
  for Computing Machinery, 2021, p. 80–88.

\bibitem{10601534}
Z.~Han, R.~Zhou, C.~Xu, Y.~Zeng, and R.~Zhang, ``Inss: An intelligent
  scheduling orchestrator for multi-gpu inference with spatio-temporal
  sharing,'' \emph{IEEE Transactions on Parallel and Distributed Systems},
  vol.~35, no.~10, pp. 1735--1748, 2024.

\bibitem{hpdc-sched}
L.~Ma, H.~Chen, E.~Shao, L.~Wang, Q.~Chen, and G.~Tan, ``Elasticroom:
  Multi-tenant dnn inference engine via co-design with resource-constrained
  compilation and strong priority scheduling,'' in \emph{Proceedings of the
  33rd International Symposium on High-Performance Parallel and Distributed
  Computing}, ser. HPDC '24.\hskip 1em plus 0.5em minus 0.4em\relax Association
  for Computing Machinery, 2024, p. 1–14.

\bibitem{multi-modal1}
B.~Hu, L.~Xu, J.~Moon, N.~J. Yadwadkar, and A.~Akella, ``Mosel: Inference
  serving using dynamic modality selection,'' 2023.

\bibitem{sigmetrics2}
A.~Kumar, A.~Sivasubramaniam, and T.~Zhu, ``Splitrpc: A {Control + Data} path
  splitting rpc stack for ml inference serving,'' \emph{Proc. ACM Meas. Anal.
  Comput. Syst.}, vol.~7, no.~2, May 2023.

\bibitem{sommelier}
P.~Guo, B.~Hu, and W.~Hu, ``Sommelier: Curating dnn models for the masses,'' in
  \emph{Proceedings of the 2022 International Conference on Management of
  Data}, ser. SIGMOD '22.\hskip 1em plus 0.5em minus 0.4em\relax New York, NY,
  USA: Association for Computing Machinery, 2022, p. 1876–1890.

\bibitem{lao2024hawkvisionlowlatencymodelessedge}
\BIBentryALTinterwordspacing
C.~Lao, J.~Gao, G.~Ananthanarayanan, A.~Akella, and M.~Yu, ``Hawkvision:
  Low-latency modeless edge ai serving,'' 2024. [Online]. Available:
  \url{https://arxiv.org/abs/2405.19213}
\BIBentrySTDinterwordspacing

\bibitem{edge-adaptor}
K.~Zhao \emph{et~al.}, ``Edgeadaptor: Online configuration adaption, model
  selection and resource provisioning for edge dnn inference serving at
  scale,'' \emph{IEEE Transactions on Mobile Computing}, vol.~22, no.~10, pp.
  5870--5886, 2023.

\bibitem{edge-inference-scheduling}
Y.~She \emph{et~al.}, ``On-demand edge inference scheduling with accuracy and
  deadline guarantee,'' in \emph{2023 IEEE/ACM 31st International Symposium on
  Quality of Service (IWQoS)}, 2023, pp. 1--10.

\bibitem{hpdc-edge}
A.~Tzenetopoulos \emph{et~al.}, ``Seamless hw-accelerated ai serving in
  heterogeneous mec systems with ai@edge,'' in \emph{Proceedings of the 33rd
  International Symposium on High-Performance Parallel and Distributed
  Computing}, ser. HPDC '24.\hskip 1em plus 0.5em minus 0.4em\relax New York,
  NY, USA: Association for Computing Machinery, 2024, p. 377–380.

\bibitem{acc-max1}
\BIBentryALTinterwordspacing
A.~Fresa and J.~P. Champati, ``Offloading algorithms for maximizing inference
  accuracy on edge device under a time constraint,'' \emph{CoRR}, vol.
  abs/2112.11413, 2021. [Online]. Available:
  \url{https://arxiv.org/abs/2112.11413}
\BIBentrySTDinterwordspacing

\bibitem{embedded-model-selection}
V.~S. Marco, B.~Taylor, Z.~Wang, and Y.~Elkhatib, ``Optimizing deep learning
  inference on embedded systems through adaptive model selection,'' \emph{ACM
  Trans. Embed. Comput. Syst.}, vol.~19, no.~1, feb 2020.

\bibitem{edge-cloud2}
X.~Wang \emph{et~al.}, ``Dynamic dnn model selection and inference off loading
  for video analytics with edge-cloud collaboration,'' in \emph{Proceedings of
  the 32nd Workshop on Network and Operating Systems Support for Digital Audio
  and Video}.\hskip 1em plus 0.5em minus 0.4em\relax New York, NY, USA:
  Association for Computing Machinery, 2022, p. 64–70.

\bibitem{neurosurgeon}
Y.~Kang, J.~Hauswald, C.~Gao, A.~Rovinski, T.~Mudge, J.~Mars, and L.~Tang,
  ``Neurosurgeon: Collaborative intelligence between the cloud and mobile
  edge,'' in \emph{Proceedings of the Twenty-Second International Conference on
  Architectural Support for Programming Languages and Operating Systems}, ser.
  ASPLOS '17.\hskip 1em plus 0.5em minus 0.4em\relax New York, NY, USA:
  Association for Computing Machinery, 2017, p. 615–629.

\bibitem{chameleon}
J.~Jiang \emph{et~al.}, ``Chameleon: Scalable adaptation of video analytics,''
  in \emph{Proceedings of the 2018 Conference of the ACM Special Interest Group
  on Data Communication}, ser. SIGCOMM '18.\hskip 1em plus 0.5em minus
  0.4em\relax New York, NY, USA: Association for Computing Machinery, 2018, p.
  253–266.

\bibitem{ekya}
R.~Bhardwaj \emph{et~al.}, ``Ekya: Continuous learning of video analytics
  models on edge compute servers,'' in \emph{19th USENIX Symposium on Networked
  Systems Design and Implementation (NSDI 22)}.\hskip 1em plus 0.5em minus
  0.4em\relax Renton, WA: USENIX Association, Apr. 2022, pp. 119--135.

\end{thebibliography}

\section{Appendix}

\subsection{Scheduling Hardness}
\label{appendix-hardness}
We can show that specific instantiations of our optimization problem in equation \ref{np-hard-opt} are NP hard, e.g. by effectively excluding model selection from the optimization.
More specifically; let $|\mathcal{M}_{a_i}| = 1$ for all applications and define $Accuracy(m_j) = 1$ for all $m_j \in \bigcup\limits_{a} \mathcal{M}_{a}$.
In addition, define the penalty function to be a step function, i.e. $\gamma_{a_i}(d_i, e_i) = \mathbbm{1}_{d_i < e_i}$.
Then we are left with an optimization problem that effectively minimizes a unit penalty for jobs with \textit{sequence independent startup times} over an arbitrary number of application families.
This is a known NP-hard problem~\cite{siam-scheduling, springer-scheduling}, even without considering more complicated choices for $\mathcal{M}$.

\subsection{Alternative Scoring Rules}
\label{appendix-scoring-rules}

We used accuracy as the evaluation metric in the main text, but the same principle applies to other scoring rules.
For example, the weighted $F_1$ score uses $\bm{\theta}$ directly when averaging. We can also rewrite the quadratic score in a similar fashion. Let $l(i)$ be the index of the true class label associated with test data point $i$ and $p_i$ be the model-generated class probabilities for each point. Then we have:
\begin{align}
    \frac{1}{n} \sum_{i=1}^n Q(p_i, l(i))
    &= \frac{1}{n} \sum_{i=1}^n 2p_{i,l(i)} - p_i^{\intercal} p_i \\
    &= 2 \sum_{i=1}^n \frac{p_{i,l(i)}}{n} - \frac{1}{n} \sum_{i=1}^n p_i^{\intercal} p_i \\
    &= 2 \sum_{j=1}^{\nc} \theta_j \mu_p(c_j) - \frac{1}{n} \sum_{i=1}^n p_i^{\intercal} p_i
\end{align}

\noindent
where $\mu_p(c_j)$ the the average probability assigned to $c_j$ when $c_j$ is the true class label. Now the inferences from our SneakPeek models can be used directly inform this scoring function.


\end{document}